\documentclass{article}%
\usepackage{amsmath}
\usepackage{amsfonts}
\usepackage{amssymb}
\usepackage{graphicx}%
\begin{document}

\author{Sawa Manoff\\\textit{Institute for Nuclear Research and Nuclear Energy}\\\textit{\ Department of Theoretical Physics}\\\textit{\ Blvd. Tzarigradsko Chaussee 72}\\\textit{\ 1784 Sofia - Bulgaria}}
\date{\textit{E-mail address: smanov@inrne.bas.bg }}
\title{\textbf{Centrifugal (centripetal) and Coriolis' velocities and accelerations
in spaces with affine connections and metrics as models of space-time}}
\maketitle

\begin{abstract}
\textit{The notions of centrifugal (centripetal) and Coriolis' velocities and
accelerations are introduced and considered in spaces with affine connections
and metrics [}$(\overline{L}_{n},g)$-\textit{spaces] as velocities and
accelerations of flows of mass elements (particles) moving in space-time. It
is shown that these types of velocities and accelerations are generated by the
relative motions between the mass elements. They are closely related to the
kinematic characteristics of the relative velocity and relative acceleration.
The centrifugal (centripetal) velocity is found to be in connection with the
Hubble law. The centrifugal (centripetal) acceleration could be interpreted as
gravitational acceleration as it has been done in the Einstein theory of
gravitation. This fact could be used as a basis for working out of new
gravitational theories in spaces with affine connections and metrics. }


\end{abstract}

\section{Introduction}

1. The relative velocity and the relative acceleration between particles or
mass elements of a flow are important characteristics for describing the
evolution and the motion of a dynamic system. On the other side, the kinematic
characteristics related to the relative velocity and the relative acceleration
(such as deformation velocity and acceleration, shear velocity and
acceleration, rotation velocity and acceleration, and expansion velocity and
acceleration) characterize specific relative motions of particles and /or mass
elements in a flow \cite{Manoff-0}$\div$\cite{Manoff-4}. On the basis of the
links between the kinematic characteristics related to the relative velocity
and these related to the relative acceleration the evolution of a system of
particles or mass elements of a flow could be connected to the geometric
properties of the corresponding mathematical model of a space or space-time
\cite{Stephani}. Many of the notions of classical mechanics or of classical
mechanics of continuous media preserve their physical interpretation in more
comprehensive spaces than the Euclidean or Minkowskian spaces, considered as
mathematical models of space or space-time \cite{Ehlers}. On this background,
the generalizations of the notions of Coriolis' and centrifugal (centripetal)
accelerations \cite{Lammerzahl} from classical mechanics in Euclidean spaces
are worth to be investigated in spaces with affine connections and metrics
[$(\overline{L}_{n},g)$-spaces] \cite{Bishop} $\div$ \cite{Manoff-7a}.

2. Usually, the Coriolis' and centrifugal (centripetal) accelerations are
considered as apparent accelerations, generated by the non-inertial motion of
the basic vector fields determining a co-ordinate system or a frame of
reference in an Euclidean space. In Einstein's theory of gravitation (ETG)
these types of accelerations are considered to be generated by a symmetric
affine connection (Riemannian connection) compatible with the corresponding
Riemannian metric \cite{Stephani}. In both cases they are considered as
corollaries of the non-inertial motion of particles (i.e. of the motion of
particles in non-inertial co-ordinate system).

3. In the present paper the notions of Coriolis' and centrifugal (centripetal)
accelerations are considered with respect to their relations with the
geometric characteristics of the corresponding models of space or space-time.
It appears that accelerations of these types are closely related to the
kinematic characteristics of the relative velocity and of the relative
acceleration. The main idea is to be found \ out how a Coriolis' or
centrifugal (centripetal) acceleration acts on a mass element of a flow or on
a single particle during its motion in space or space-time described by a
space of affine connections and metrics. In Section 1 the notions of
centrifugal (centripetal) and Coriolis' velocities are introduced and
considered in $(\overline{L}_{n},g)$-spaces. The relations between the
kinematic characteristics of the relative velocity and the introduced notions
are established. The centrifugal (centripetal) velocity is found to be in
connection with the Hubble law in spaces with affine connections and metrics.
In Section 2 the notions of centrifugal (centripetal) and Coriolis'
accelerations are introduced and their relations to the kinematic
characteristics of the relative acceleration are found. In Section 3 the
interpretation of the centrifugal (centripetal) acceleration as gravitational
acceleration is given and illustrated on the basis of the Einstein theory of
gravitation and especially on the basis of the Schwarzschild metric in vacuum.

4. The main results in the paper are given in details (even in full details)
for these readers who are not familiar with the considered problems. The
definitions and abbreviations are identical to those used in \cite{Manoff-5},
\cite{Manoff-7}, \cite{Manoff-7a}. The reader is kindly asked to refer to them
for more details and explanations of the statements and results only cited in
this paper.

\section{Centrifugal (centripetal) and Coriolis' velocities}

Let us now recall some well known facts from kinematics of vector fields over
spaces with affine connections and metrics [$(\overline{L}_{n},g)$ - spaces],
considered as models of space or space-time \cite{Manoff-0}$\div
$\cite{Manoff-7a}.

1. Every contravariant vector field $\xi$ could be represented by the use of a
non-isotropic (non-null) contravariant vector field $u$ and its corresponding
contravariant and covariant projective metrics $h^{u}$ and $h_{u}$ in the
form
\begin{equation}
\xi=\frac1e\cdot g(u,\xi)\cdot u+\overline{g}[h_{u}(\xi)]\text{ ,} \label{1.1}%
\end{equation}
where%

\begin{align}
h^{u}  &  =\overline{g}-\frac{1}{e}\cdot u\otimes u\text{ ,~\ \ \ \ \ \ }%
h_{u}=g-\frac{1}{e}\cdot g(u)\otimes g(u)\text{ ,}\label{1.2}\\
\overline{g}  &  =g^{ij}\cdot e_{i}.e_{j}\text{ , \ \ \ }g^{ij}=g^{ji}\text{ ,
\ \ }e_{i}.e_{j}=\frac{1}{2}\cdot(e_{i}\otimes e_{j}+e_{j}\otimes e_{i})\text{
,}\nonumber\\
g  &  =g_{ij}\cdot e^{i}.e^{j}\text{ , \ \ \ \ }g_{ij}=g_{ji}\text{ \ ,
\ \ }e^{i}.e^{j}=\frac{1}{2}\cdot(e^{i}\otimes e^{j}+e^{j}\otimes e^{i})\text{
,\ }\nonumber
\end{align}

\begin{align}
e  &  =g(u,u)=g_{\overline{k}\overline{l}}\cdot u^{k}\cdot u^{l}%
=u_{\overline{k}}\cdot u^{k}=u_{k}\cdot u^{\overline{k}}:\neq0\text{ ,
\ }\label{1.2a}\\
g(u,\xi)  &  =g_{\overline{i}\overline{j}}\cdot u^{i}\cdot\xi^{j}\text{ ,
}e_{i}=\partial_{i}\text{ \ , \ \ \ }e^{j}=dx^{j}\text{ \ in a co-ordinate
basis.} \label{1.2b}%
\end{align}

By means of the representation of $\xi$ the notions of relative velocity and
relative acceleration are introduced \cite{Manoff-5}, \cite{Manoff-7} in
$(\overline{L}_{n},g)$ - spaces. By that, the contravariant vector field $\xi$
has been considered as vector field orthogonal to the contravariant vector
field $u$, i.e. $g(u,\xi)=0$, $\xi=\xi_{\perp}=\overline{g}[h_{u}(\xi)]$. Both
the fields are considered as tangent vector fields to the corresponding
co-ordinates, i.e. they fulfil the condition \cite{Bishop} $\pounds _{\xi
}u=-\pounds _{u}\xi=[u,\xi]=0$, where $[u,\xi]=u\circ\xi-\xi\circ u$. Under
these preliminary conditions the relative velocity $_{rel}v$ could be found in
the form \cite{Manoff-5}, \cite{Manoff-7}
\begin{equation}
_{rel}v=\overline{g}[d(\xi_{\perp})]\text{ \ ,} \label{1.3}%
\end{equation}
where
\begin{equation}
d=\sigma+\omega+\frac1{n-1}\cdot\theta\cdot h_{u}\text{ \ .} \label{1.4}%
\end{equation}

The covariant tensor $d$ is the deformation velocity tensor; the covariant
symmetric trace-free [$\overline{g}[\sigma]=g^{\overline{i}\overline{j}}%
\cdot\sigma_{ij}=0$] tensor $\sigma$ is the shear velocity tensor; the
covariant antisymmetric tensor $\omega$ is the rotation velocity tensor; the
invariant $\theta$ is the expansion velocity invariant. The contravariant
vector field $_{rel}v$ is interpreted as the relative velocity of two points
(mass elements, particles) moving in a space or space-time and having equal
proper times \cite{Manoff-0}$\div$\cite{Manoff-4}. The vector field
$\xi_{\perp}$ is orthogonal to $u$ and is interpreted as deviation vector
connecting the two mass elements (particles) (if considered as an
infinitesimal vector field).

2. Let us now consider the representation of the relative velocity by the use
of a non-isotropic (non-null) contravariant vector field $\xi_{\perp}$
(orthogonal to $u$) and its corresponding projective metrics
\begin{align}
h^{\xi_{\perp}}  &  =\overline{g}-\frac1{g(\xi_{\perp},\xi_{\perp})}\cdot
\xi_{\perp}\otimes\xi_{\perp}\text{ \ ,}\label{1.5}\\
h_{\xi_{\perp}}  &  =g-\frac1{g(\xi_{\perp},\xi_{\perp})}\cdot g(\xi_{\perp
})\otimes g(\xi_{\perp})\text{ \ .} \label{1.6}%
\end{align}

Then a contravariant vector field $v$ could be represented in the form
\begin{equation}
v=\frac{g(v,\xi_{\perp})}{g(\xi_{\perp},\xi_{\perp})}\cdot\xi_{\perp
}+\overline{g}[h_{\xi_{\perp}}(v)]\text{ \ .} \label{1.7}%
\end{equation}

Therefore, the relative velocity $_{rel}v$ as a contravariant vector field
could be now written in the form
\begin{equation}
_{rel}v=\frac{g(_{rel}v,\xi_{\perp})}{g(\xi_{\perp},\xi_{\perp})}\cdot
\xi_{\perp}+\overline{g}[h_{\xi_{\perp}}(_{rel}v)]=v_{z}+v_{c}\text{ \ \ ,}
\label{1.8}%
\end{equation}
where
\begin{equation}
v_{z}=\frac{g(_{rel}v,\xi_{\perp})}{g(\xi_{\perp},\xi_{\perp})}\cdot\xi
_{\perp}\text{ \ , \ \ \ \ \ \ \ }v_{c}=\overline{g}[h_{\xi_{\perp}}%
(_{rel}v)]\text{ \ .} \label{1.9}%
\end{equation}

The vector field $v_{z}$ is collinear to the vector field $\xi_{\perp}$. If
the factor (invariant) before $\xi_{\perp}$ is positive, i.e. if
\begin{equation}
\frac{g(_{rel}v,\xi_{\perp})}{g(\xi_{\perp},\xi_{\perp})}>0\text{ }
\label{1.10}%
\end{equation}
the vector field $v_{z}$ is called (relative) \textit{centrifugal velocity}.
If the factor (invariant) before $\xi_{\perp}$ is negative, i.e. if
\begin{equation}
\frac{g(_{rel}v,\xi_{\perp})}{g(\xi_{\perp},\xi_{\perp})}<0 \label{1.11}%
\end{equation}
the vector field $v_{z}$ is called (relative) \textit{centripetal velocity}.

The vector field $v_{c}$ is called (relative) \textit{Coriolis' velocity.}

\textit{Remark}. The word (relative) is used here to stress the fact that the
centripetal and the Coriolis velocities appear with respect to a trajectory
interpreted as the world line of an observer with his sub space $T^{\perp
u}(M)$, orthogonal to $u$ at every point of his world line..

3. We can introduce the (normalized) contravariant unit vectors $n_{\parallel
}$ and $n_{\perp}$ as
\begin{equation}
n_{\parallel}=\frac1{l_{u}}\cdot u\text{ \thinspace\thinspace\thinspace
\thinspace\thinspace,\thinspace\thinspace\thinspace\thinspace\thinspace
\thinspace\thinspace\thinspace\thinspace\thinspace\thinspace\thinspace
\thinspace\thinspace\thinspace\thinspace\thinspace}n_{\perp}=\frac
1{l_{\xi_{\perp}}}\cdot\xi_{\perp}\,\,\,\,\text{,} \label{1.11a}%
\end{equation}
where
\begin{align}
g(u,u)  &  =e=\pm l_{u}^{2}\,\,\,\,\,\text{,\thinspace\thinspace
\thinspace\thinspace\thinspace\thinspace\thinspace\thinspace\thinspace
\thinspace\thinspace\thinspace\thinspace}g(\xi_{\perp},\xi_{\perp})=\mp
l_{\xi_{\perp}}^{2}\,\,\,\,\text{,}\label{1.11b}\\
g(n_{\parallel},n_{\parallel})  &  =\frac1{l_{u}^{2}}\cdot g(u,u)=\pm
\frac{l_{u}^{2}}{l_{u}^{2}}=\pm1\,\,\,\text{,}\label{1.11c}\\
g(n_{\perp},n_{\perp})  &  =\frac1{l_{\xi_{\perp}}^{2}}\cdot g(\xi_{\perp}%
,\xi_{\perp})=\mp\frac{l_{\xi_{\perp}}^{2}}{l_{\xi_{\perp}}^{2}}%
=\mp1\,\,\,\,\text{.} \label{1.11d}%
\end{align}

\textit{Remark}. The signs $(\pm)$ and $(\mp)$ are depending on the signature
of the metric $g$ of the space or space-time. If $Sgn\,g>0$ then
$g(u,u)=-l_{u}^{2}$, $g(\xi_{\perp},\xi_{\perp})=+l_{\xi_{\perp}}^{2}$,
$g(n_{\parallel},n_{\parallel})=-1$, $g(n_{\perp},n_{\perp})=+1$. If
$Sgn\,g<0$ then $g(u,u)=+l_{u}^{2}$, $g(\xi_{\perp},\xi_{\perp})=-l_{\xi
_{\perp}}^{2}$, $g(n_{\parallel},n_{\parallel})=+1$, $g(n_{\perp},n_{\perp
})=-1$.

By the use of the contravariant unit vectors $n_{\parallel}$ and $n_{\perp} $
the (relative) centrifugal (centripetal) velocity $v_{z}$ could be written in
the form
\begin{equation}
v_{z}=\mp g(_{rel}v,n_{\perp})\cdot n_{\perp}\,\,\,\text{.} \label{1.11e}%
\end{equation}

\subsection{Centrifugal (centripetal) velocity}

\paragraph{Properties of the centrifugal (centripetal) velocity}

(a) Since $v_{z}$ is collinear to $\xi_{\perp}$, it is orthogonal to the
vector field $u$, i.e.
\begin{equation}
g(u,v_{z})=0\text{ \ .} \label{1.12}%
\end{equation}

(b) The centrifugal (centripetal) velocity $v_{z}$ is orthogonal to the
Coriolis velocity $v_{c}$%
\begin{equation}
g(v_{z},v_{c})=0\text{ .} \label{1.13}%
\end{equation}

(c) The length of the vector $v_{z}$ could be found by means of the relation
\begin{align}
g(v_{z},v_{z})  &  =v_{z}^{2}=\mp l_{v_{z}}^{2}=g(\frac{g(_{rel}v,\xi_{\perp
})}{g(\xi_{\perp},\xi_{\perp})}\cdot\xi_{\perp},\frac{g(_{rel}v,\xi_{\perp}%
)}{g(\xi_{\perp},\xi_{\perp})}\cdot\xi_{\perp})=\label{1.14}\\
&  =\frac{[g(_{rel}v,\xi_{\perp})]^{2}}{g(\xi_{\perp},\xi_{\perp})^{2}}\cdot
g(\xi_{\perp},\xi_{\perp})=\frac{[g(_{rel}v,\xi_{\perp})]^{2}}{g(\xi_{\perp
},\xi_{\perp})}=\frac{[g(_{rel}v,\xi_{\perp})]^{2}}{\xi_{\perp}^{2}%
}=\nonumber\\
&  =\mp\frac{[g(_{rel}v,\xi_{\perp})]^{2}}{l_{\xi_{\perp}}^{2}}=\mp
\frac{l_{\xi_{\perp}}^{2}\cdot[g(_{rel}v,n_{\perp})]^{2}}{l_{\xi_{\perp}}^{2}%
}=\mp[g(_{rel}v,n_{\perp})]^{2}\,\,\text{.} \label{1.14a}%
\end{align}

From the last (previous) expression we can conclude that the square of the
length of the centrifugal (centripetal) velocity is in general in inverse
proportion to the length of the vector field $\xi_{\perp}$.

\textit{Special case:} $M_{n}:=E_{n}$, $n=3$ (3-dimensional Euclidean space):
$\xi_{\perp}=\overrightarrow{r}$, $g(\xi_{\perp},\xi_{\perp})=r^{2}$%
\begin{equation}
v_{z}^{2}=\frac{[g(\overrightarrow{r},_{rel}\overrightarrow{v})]^{2}}{r^{2}%
}\text{ , \thinspace\thinspace\thinspace\thinspace\thinspace\thinspace
\thinspace\thinspace\thinspace\thinspace\thinspace\thinspace\thinspace
\thinspace\thinspace\thinspace}l_{v_{z}}=\sqrt{v_{z}^{2}}=\frac
{g(\overrightarrow{r},_{rel}\overrightarrow{v})}r\text{ \thinspace
\thinspace\thinspace\thinspace.} \label{1.15}%
\end{equation}

If the relative velocity $_{rel}v$ is equal to zero then $v_{z}=0$.

(d) The scalar product between $v_{z}$ and $_{rel}v$ could be found in its
explicit form by the use of the explicit form of the relative velocity
$_{rel}v$%
\begin{equation}
_{rel}v=\overline{g}[d(\xi_{\perp})]=\overline{g}[\sigma(\xi_{\perp
})]+\overline{g}[\omega(\xi_{\perp})]+\frac{1}{n-1}\cdot\theta\cdot\xi_{\perp
}\text{ ,} \label{1.16}%
\end{equation}
and the relations
\begin{equation}
g(\xi_{\perp},_{rel}v)=g(\xi_{\perp},\overline{g}[\sigma(\xi_{\perp
})])+\overline{g}(\xi_{\perp},\overline{g}[\omega(\xi_{\perp})])+\frac{1}%
{n-1}\cdot\theta\cdot g(\xi_{\perp},\xi_{\perp})\text{ ,} \label{1.17}%
\end{equation}
\begin{align}
g(\xi_{\perp},\overline{g}[\sigma(\xi_{\perp})])  &  =\sigma(\xi_{\perp}%
,\xi_{\perp})\text{ ,}\label{1.18}\\
\overline{g}(\xi_{\perp},\overline{g}[\omega(\xi_{\perp})])  &  =\omega
(\xi_{\perp},\xi_{\perp})=0\text{ ,}\label{1.19}\\
g(\xi_{\perp},_{rel}v)  &  =\sigma(\xi_{\perp},\xi_{\perp})+\frac{1}{n-1}%
\cdot\theta\cdot g(\xi_{\perp},\xi_{\perp})\text{ .} \label{1.20}%
\end{align}

For
\begin{align}
g(v_{z},_{rel}v)  &  =g(\frac{g(\xi_{\perp},_{rel}v)}{g(\xi_{\perp},\xi
_{\perp})}.\xi_{\perp},_{rel}v)=\frac{g(\xi_{\perp},_{rel}v)}{g(\xi_{\perp
},\xi_{\perp})}\cdot g(\xi_{\perp},_{rel}v)=\nonumber\\
&  =\frac{[g(\xi_{\perp},_{rel}v)]^{2}}{g(\xi_{\perp},\xi_{\perp})}=v_{z}%
^{2}\text{ ,} \label{1.21}%
\end{align}
it follows that
\begin{equation}
g(v_{z},_{rel}v)=g(v_{z},v_{z})=v_{z}^{2}\text{ .} \label{1.21a}%
\end{equation}

\textit{Remark.} $g(v_{z},_{rel}v)=g(v_{z},v_{z})=v_{z}^{2}$ because of
$g(v_{z},v_{c})=0$ and $g(v_{z},_{rel}v)=g(v_{z},v_{z}+v_{c})=g(v_{z}%
,v_{z})+g(v_{z},v_{c})=g(v_{z},v_{z})$.

On the other side, we can express $v_{z}^{2}$ by the use of the kinematic
characteristics of the relative velocity $\sigma$, $\omega$, and $\theta$%
\begin{align}
v_{z}^{2}  &  =\frac{[g(\xi_{\perp},_{rel}v)]^{2}}{g(\xi_{\perp},\xi_{\perp}%
)}=\frac1{g(\xi_{\perp},\xi_{\perp})}\cdot[\sigma(\xi_{\perp},\xi_{\perp
})+\frac1{n-1}\cdot\theta\cdot g(\xi_{\perp},\xi_{\perp})]^{2}=\nonumber\\
&  =\frac1{g(\xi_{\perp},\xi_{\perp})}\cdot\{[\sigma(\xi_{\perp},\xi_{\perp
})]^{2}+\frac1{(n-1)^{2}}\cdot\theta^{2}\cdot[g(\xi_{\perp},\xi_{\perp}%
)]^{2}+\nonumber\\
&  +\frac2{n-1}\cdot\theta\cdot\sigma(\xi_{\perp},\xi_{\perp})\cdot
g(\xi_{\perp},\xi_{\perp})\}\text{ ,}\nonumber\\
v_{z}^{2}  &  =\frac{[\sigma(\xi_{\perp},\xi_{\perp})]^{2}}{g(\xi_{\perp}%
,\xi_{\perp})}+\frac1{(n-1)^{2}}\cdot\theta^{2}\cdot g(\xi_{\perp},\xi_{\perp
})+\frac2{n-1}\cdot\theta\cdot\sigma(\xi_{\perp},\xi_{\perp})\text{ ,}
\label{1.22}%
\end{align}
\begin{align*}
v_{z}^{2}  &  =\mp l_{v_{z}}^{2}=\mp l_{\xi_{\perp}}^{2}\cdot[\sigma(n_{\perp
},n_{\perp})]^{2}\mp\frac1{(n-1)^{2}}\cdot\theta^{2}\cdot l_{\xi_{\perp}}%
^{2}+\frac2{n-1}\cdot\theta\cdot\xi_{\perp}^{2}\cdot\sigma(n_{\perp},n_{\perp
})=\\
&  =\mp l_{\xi_{\perp}}^{2}\cdot\{[\sigma(n_{\perp},n_{\perp})]^{2}%
+\frac1{(n-1)^{2}}\cdot\theta^{2}\mp\frac2{n-1}\cdot\theta\cdot\sigma
(n_{\perp},n_{\perp})\}=\\
&  =\mp l_{\xi_{\perp}}^{2}\cdot[\sigma(n_{\perp},n_{\perp})\mp\frac
1{n-1}\cdot\theta]^{2}\text{ \thinspace\thinspace\thinspace,}%
\end{align*}
\begin{equation}
l_{v_{z}}^{2}=l_{\xi_{\perp}}^{2}\cdot[\sigma(n_{\perp},n_{\perp})\mp
\frac1{n-1}\cdot\theta]^{2}=[\frac1{n-1}\cdot\theta\mp\sigma(n_{\perp
},n_{\perp})]^{2}\cdot l_{\xi_{\perp}}^{2}\,\,\,\text{.} \label{1.22a}%
\end{equation}

Therefore, for $l_{v_{z}}$ we obtain
\begin{align}
l_{v_{z}}  &  =\pm[\sigma(n_{\perp},n_{\perp})\mp\frac1{n-1}\cdot\theta]\cdot
l_{\xi_{\perp}}\,=\label{1.22b}\\
&  =\,\,\,[\frac1{n-1}\cdot\theta\mp\sigma(n_{\perp},n_{\perp})]\cdot
l_{\xi_{\perp}}\,\,\,\text{.}%
\end{align}

\textit{Special case:} $\theta:=0$ (expansion-free relative velocity)
\begin{equation}
v_{z}^{2}=\frac{[\sigma(\xi_{\perp},\xi_{\perp})]^{2}}{g(\xi_{\perp}%
,\xi_{\perp})}=\frac{[\sigma(\xi_{\perp},\xi_{\perp})]^{2}}{\xi_{\perp}^{2}%
}=\frac{[\sigma(\xi_{\perp},\xi_{\perp})]^{2}}{\mp l_{\xi_{\perp}}^{2}}\text{
,} \label{1.23}%
\end{equation}
\begin{align}
g(\xi_{\perp},\xi_{\perp})  &  =\mp l_{\xi_{\perp}}^{2}\text{ \thinspace
\thinspace,\thinspace\thinspace\thinspace\thinspace}\label{1.23a}\\
g(\xi_{\perp},\xi_{\perp})  &  =-l_{\xi_{\perp}}^{2}\text{\thinspace
\thinspace}\,\,\,\text{if \thinspace\thinspace\thinspace\thinspace
\thinspace\thinspace}g(u,u)=e=+l_{u}^{2}\text{ \thinspace\thinspace
,}\label{1.23b}\\
g(\xi_{\perp},\xi_{\perp})  &  =+l_{\xi_{\perp}}^{2}\text{\thinspace
}\,\,\,\,\,\text{if\thinspace\thinspace\thinspace\thinspace\thinspace
\thinspace\thinspace\thinspace}g(u,u)=e=-l_{u}^{2}\text{ \thinspace
\thinspace\thinspace.\thinspace\thinspace\thinspace\thinspace} \label{1.23c}%
\end{align}

\textit{Remark}. If we introduce as above the unit vector field $n_{\perp}$,
normal to $u$ and normalized, as
\begin{align}
n_{\perp}  &  :=\frac{\xi_{\perp}}{l_{\xi_{\perp}}}\text{ \ , \thinspace
\ \ \ \ }g(\xi_{\perp},\xi_{\perp})=\mp l_{\xi_{\perp}}^{2}\text{ \ ,
\thinspace\thinspace\thinspace\thinspace}\xi_{\perp}=l_{\xi_{\perp}}\cdot
n_{\perp}\text{ \thinspace\thinspace\thinspace\thinspace,\ \ \ \ \ \ \ }%
\label{1.28a}\\
l_{\xi_{\perp}}  &  =\,\mid g(\xi_{\perp},\xi_{\perp})\mid^{1/2}\text{ ,
\ \ \ }g(n_{\perp},n_{\perp})=\mp1\text{ \ .} \label{1.28b}%
\end{align}
then the above expressions for $v_{z}^{2}$ with $[\sigma(\xi_{\perp}%
,\xi_{\perp})]^{2}/$ $g(\xi_{\perp},\xi_{\perp})$ could be written in the
form:
\begin{equation}
v_{z}^{2}=\frac{[\sigma(\xi_{\perp},\xi_{\perp})]^{2}}{\mp l_{\xi_{\perp}}%
^{2}}=\mp l_{\xi_{\perp}}^{2}\cdot[\sigma(n_{\perp},n_{\perp})]^{2}%
\,\,\,\,\,\text{.} \label{1.28d}%
\end{equation}

\textit{Special case: }$\sigma:=0$ (shear-free relative velocity)
\begin{align}
v_{z}^{2}  &  =\mp l_{v_{z}}^{2}=\frac1{(n-1)^{2}}\cdot\theta^{2}\cdot
g(\xi_{\perp},\xi_{\perp})=\mp\frac1{(n-1)^{2}}\cdot\theta^{2}\cdot
l_{\xi_{\perp}}^{2}\text{ ,}\label{1.23d}\\
l_{v_{z}}^{2}  &  =\frac1{(n-1)^{2}}\cdot\theta^{2}\cdot l_{\xi_{\perp}}%
^{2}\,\,\,\text{.} \label{1.23e}%
\end{align}

(e) The explicit form of $v_{z}$ could be found as
\begin{align}
v_{z}  &  =\frac{g(_{rel}v,\xi_{\perp})}{g(\xi_{\perp},\xi_{\perp})}\cdot
\xi_{\perp}=\frac1{g(\xi_{\perp},\xi_{\perp})}\cdot[\sigma(\xi_{\perp}%
,\xi_{\perp})+\frac1{n-1}\cdot\theta\cdot g(\xi_{\perp},\xi_{\perp})]\cdot
\xi_{\perp}=\nonumber\\
&  =[\frac{\sigma(\xi_{\perp},\xi_{\perp})}{g(\xi_{\perp},\xi_{\perp})}%
+\frac1{n-1}\cdot\theta]\cdot\xi_{\perp}=[\frac1{n-1}\cdot\theta\mp
\sigma(n_{\perp},n_{\perp})]\cdot\xi_{\perp}\text{ ,}\nonumber\\
v_{z}  &  =[\frac1{n-1}\cdot\theta+\frac{\sigma(\xi_{\perp},\xi_{\perp}%
)}{g(\xi_{\perp},\xi_{\perp})}]\cdot\xi_{\perp}=[\frac1{n-1}\cdot\theta
\mp\sigma(n_{\perp},n_{\perp})]\cdot l_{\xi_{\perp}}\cdot n_{\perp}\text{ .}
\label{1.24}%
\end{align}

If
\begin{align}
\frac1{n-1}\cdot\theta+\frac{\sigma(\xi_{\perp},\xi_{\perp})}{g(\xi_{\perp
},\xi_{\perp})}  &  >0\text{ , }\label{1.25}\\
\frac1{n-1}\cdot\theta\mp\sigma(n_{\perp},n_{\perp})  &
>0\,\,\,\,\text{,\thinspace\thinspace\thinspace\thinspace\thinspace
\thinspace\thinspace\thinspace\thinspace\ }\xi_{\perp}^{2}=g(\xi_{\perp}%
,\xi_{\perp})=\mp l_{\xi_{\perp}}^{2}\text{ ,}%
\end{align}
we have a centrifugal (relative) velocity.

If
\begin{align}
\frac1{n-1}\cdot\theta+\frac{\sigma(\xi_{\perp},\xi_{\perp})}{g(\xi_{\perp
},\xi_{\perp})}  &  <0\text{ ,}\label{1.26}\\
\frac1{n-1}\cdot\theta\mp\sigma(n_{\perp},n_{\perp})  &  <0\text{ ,}%
\end{align}
we have a centripetal (relative) velocity.

\textit{Special case: }$\theta:=0$ (expansion-free relative velocity)
\begin{equation}
v_{z}=\frac{\sigma(\xi_{\perp},\xi_{\perp})}{g(\xi_{\perp},\xi_{\perp})}%
\cdot\xi_{\perp}=\mp\sigma(n_{\perp},n_{\perp})\cdot l_{\xi_{\perp}}\cdot
n_{\perp}\text{ .} \label{1.26a}%
\end{equation}

\textit{Special case:} $\sigma:=0$ (shear-free relative velocity)
\begin{equation}
v_{z}=\frac1{n-1}\cdot\theta\cdot\xi_{\perp}=\frac1{n-1}\cdot\theta\cdot
l_{\xi_{\perp}}\cdot n_{\perp}\text{ \ .} \label{1.27}%
\end{equation}

If the expansion invariant $\theta>0$ we have centrifugal (or expansion)
(relative) velocity. If the expansion invariant $\theta<0$ we have centripetal
(or contraction) (relative) velocity. Therefore, in the case of a shear-free
relative velocity $v_{z}$ is proportional to the expansion velocity invariant
$\theta$.

The centrifugal (centripetal) relative velocity $v_{z}$ could be also written
in the form
\begin{align}
v_{z}  &  =\frac{g(_{rel}v,\xi_{\perp)}}{g(\xi_{\perp},\xi_{\perp})}\cdot
\xi_{\perp}=\mp\frac{g(_{rel}v,\xi_{\perp)}}{l_{\xi_{\perp}}^{2}}\cdot
\xi_{\perp}=\nonumber\\
&  =\mp g(_{rel}v,\frac{\xi_{\perp}}{l_{\xi_{\perp}}})\cdot\frac{\xi_{\perp}%
}{l_{\xi_{\perp}}}=\mp g(_{rel}v,n_{\perp})\cdot n_{\perp}\text{ .}
\label{1.28e}%
\end{align}

On the other side, on the basis of the relations
\begin{align*}
_{rel}v  &  =\overline{g}[d(\xi_{\perp})]=\overline{g}[\sigma(\xi_{\perp
})]+\overline{g}[\omega(\xi_{\perp})]+\frac1{n-1}\cdot\theta\cdot\xi_{\perp
}\text{ ,}\\
\overline{g}[h_{u}(\xi)]  &  =\xi_{\perp}\text{ \thinspace\thinspace
\thinspace,\thinspace\thinspace\thinspace\thinspace\thinspace\thinspace
\thinspace}\overline{g}[h_{u}(\xi_{\perp})]=\overline{g}[h_{u}(\overline
{g}[h_{u}(\xi)])]=\overline{g}[h_{u}(\xi)]=\xi_{\perp}\text{ ,}\\
g(\overline{g}[\sigma(\xi_{\perp})],n_{\perp})  &  =\sigma(n_{\perp}%
,\xi_{\perp})=\sigma(n_{\perp},l_{\xi_{\perp}}\cdot n_{\perp})=l_{\xi_{\perp}%
}\cdot\sigma(n_{\perp},n_{\perp})\text{ ,}\\
g(\overline{g}[\omega(\xi_{\perp})],n_{\perp})  &  =\omega(n_{\perp}%
,\xi_{\perp})=\omega(n_{\perp},l_{\xi_{\perp}}\cdot n_{\perp})=l_{\xi_{\perp}%
}\cdot\omega(n_{\perp},n_{\perp})=0\text{ \thinspace\thinspace,}\\
g(\xi_{\perp},n_{\perp})  &  =l_{\xi_{\perp}}\cdot g(n_{\perp},n_{\perp})=\mp
l_{\xi_{\perp}}\text{ ,}%
\end{align*}
it follows for $g(_{rel}v,n_{\perp})$%
\begin{equation}
g(_{rel}v,n_{\perp})=[\sigma(n_{\perp},n_{\perp})\mp\frac1{n-1}\cdot
\theta]\cdot l_{\xi_{\perp}}\text{ \thinspace\thinspace,} \label{1.28f}%
\end{equation}
and for $v_{z}$%
\begin{align}
v_{z}  &  =\mp g(_{rel}v,n_{\perp})\cdot n_{\perp}=[\mp\sigma(n_{\perp
},n_{\perp})+\frac1{n-1}\cdot\theta]\cdot l_{\xi_{\perp}}\cdot n_{\perp}\text{
\thinspace\thinspace,}\nonumber\\
v_{z}  &  =[\frac1{n-1}\cdot\theta\mp\sigma(n_{\perp},n_{\perp})]\cdot
l_{\xi_{\perp}}\cdot n_{\perp}\text{ \thinspace\thinspace.} \label{1.28g}%
\end{align}

Since $l_{\xi_{\perp}}>0$, we have three different cases:

(a)
\begin{align*}
v_{z}  &  >0:\mp\sigma(n_{\perp},n_{\perp})+\frac1{n-1}\cdot\theta>0:\\
\theta &  >\pm(n-1)\cdot\sigma(n_{\perp},n_{\perp})\text{ ,\thinspace
\thinspace\thinspace\thinspace\thinspace\thinspace\thinspace\thinspace
\thinspace\thinspace\thinspace}n-1>0\text{ .}%
\end{align*}

(b)
\begin{align*}
v_{z}  &  <0:\mp\sigma(n_{\perp},n_{\perp})+\frac1{n-1}\cdot\theta<0:\\
\theta &  <\pm(n-1)\cdot\sigma(n_{\perp},n_{\perp})\text{ ,\thinspace
\thinspace\thinspace\thinspace\thinspace\thinspace\thinspace\thinspace
\thinspace\thinspace\thinspace}n-1>0\text{ .}%
\end{align*}

(c)
\begin{align*}
v_{z}  &  =0:\mp\sigma(n_{\perp},n_{\perp})+\frac1{n-1}\cdot\theta=0:\\
\theta &  =\pm(n-1)\cdot\sigma(n_{\perp},n_{\perp})\text{ ,\thinspace
\thinspace\thinspace\thinspace\thinspace\thinspace\thinspace\thinspace
\thinspace\thinspace\thinspace}n-1>0\text{ \thinspace\thinspace\thinspace
\thinspace\thinspace.}%
\end{align*}

\textit{Special case}: $\sigma:=0$ (shear-free relative velocity)
\[
v_{z}=\frac1{n-1}\cdot\theta\cdot l_{\xi_{\perp}}\cdot n_{\perp}\text{
\thinspace\thinspace.}
\]

For $v_{z}>0:\theta>0$ we have an expansion, for $v_{z}<0:\theta<0$ we have a
contraction, and for $v_{z}=0:\theta=0$ we have a stationary case.

\textit{Special case:} $\theta:=0$ (expansion-free relative velocity)
\[
v_{z}=\mp\sigma(n_{\perp},n_{\perp})\cdot l_{\xi_{\perp}}\cdot n_{\perp}\text{
\thinspace\thinspace.}
\]

For $v_{z}>0:\mp\sigma(n_{\perp},n_{\perp})>0$ we have an expansion, for
$v_{z}<0:\mp\sigma(n_{\perp},n_{\perp})<0$ we have a contraction, and for
$v_{z}=0:\sigma(n_{\perp},n_{\perp})=0$ we have a stationary case.

In an analogous way we can find the explicit form of $v_{z}^{2}$ as
\begin{align}
v_{z}^{2}  &  =\{\mp[\sigma(n_{\perp},n_{\perp})]^{2}\mp\frac1{(n-1)^{2}}%
\cdot\theta^{2}+\frac2{n-1}\cdot\theta\cdot\sigma(n_{\perp},n_{\perp})\}\cdot
l_{\xi_{\perp}}^{2}=\label{1.28h}\\
&  =\mp[\sigma(n_{\perp},n_{\perp})\mp\frac1{n-1}\cdot\theta]^{2}\cdot
l_{\xi_{\perp}}^{2}\text{ .} \label{1.28i}%
\end{align}

The expression for $v_{z}^{2}$ could be now written in the form
\begin{equation}
v_{z}^{2}=\mp H^{2}\cdot l_{\xi_{\perp}}^{2}=\mp l_{v_{z}}^{2}\text{\thinspace
\thinspace\thinspace\thinspace\thinspace\thinspace\thinspace,} \label{1.28j}%
\end{equation}
where
\begin{equation}
H^{2}=[\sigma(n_{\perp},n_{\perp})\mp\frac1{n-1}\cdot\theta]^{2}\text{ .}
\label{1.28k}%
\end{equation}

Then
\begin{align}
l_{v_{z}}^{2}  &  =H^{2}\cdot l_{\xi_{\perp}}^{2}\text{ \thinspace
\thinspace\thinspace\thinspace,\thinspace\thinspace\thinspace\thinspace
\thinspace\thinspace\thinspace\thinspace\thinspace\thinspace\thinspace
\thinspace}l_{v_{z}}=\,\mid H\mid\cdot l_{\xi_{\perp}}\text{\thinspace
\thinspace\thinspace\thinspace\thinspace,\thinspace\thinspace\thinspace
\thinspace\thinspace\thinspace}l_{v_{z}}>0\text{ \thinspace\thinspace
,\thinspace\thinspace\thinspace\thinspace\thinspace}l_{\xi_{\perp}%
}>0\,\,\,\,\text{,}\label{1.28l}\\
H  &  =\mp[\sigma(n_{\perp},n_{\perp})\mp\frac1{n-1}\cdot\theta]=
\label{1.28m}\\
&  =\frac1{n-1}\cdot\theta\mp\sigma(n_{\perp},n_{\perp})\,\,\,\,\,\,\text{.}
\label{1.28n}%
\end{align}
and
\begin{equation}
v_{z}=\pm\,l_{v_{z}}\cdot n_{\perp}=\,\pm\,\mid H\mid\cdot l_{\xi_{\perp}%
}\cdot n_{\perp}=H\cdot l_{\xi_{\perp}}\cdot n_{\perp}=\,H\cdot\xi_{\perp
}\text{ \thinspace\thinspace\thinspace.} \label{1.28o}%
\end{equation}

\textit{Remark}. From the forms of $v_{z}=\pm l_{v_{z}}\cdot n_{\perp}$ and
$v_{z}=$ $[\frac1{n-1}\cdot\theta\mp\sigma(n_{\perp},n_{\perp})]\cdot
l_{\xi_{\perp}}\cdot n_{\perp}$ it follows immediately that
\begin{equation}
\pm l_{v_{z}}=H\cdot l_{\xi_{\perp}}\,\,\,\,\,\text{, \thinspace
\thinspace\thinspace\thinspace\thinspace\thinspace\thinspace\thinspace
\thinspace\thinspace\thinspace}v_{z}=H\cdot l_{\xi_{\perp}}\cdot n_{\perp
}\,\,\,\,\,\text{.} \label{1.28s}%
\end{equation}

It follows from the last (previously) expression that the centrifugal
(centripetal) relative velocity $v_{z}=H\cdot\xi_{\perp}$ is collinear to the
vector field $\xi_{\perp}$.

The absolute value $l_{v_{z}}=\,\mid H\mid\cdot l_{\xi_{\perp}}$ of $v_{z}$
and the expression for the centrifugal (centripetal) relative velocity
$v_{z}=H\cdot l_{\xi_{\perp}}\cdot n_{\perp}=\,H\cdot\xi_{\perp}$ represent
generalizations of the \textit{Hubble law }\cite{Misner}, \cite{Stephani}. The
function $H$ could be called \textit{Hubble function }(some authors call it
\textit{Hubble coefficient} \cite{Misner}, \cite{Stephani}). Since
$H=H(x^{k}(\tau))=H(\tau)$, for a given proper time $\tau=\tau_{0}$ the
function $H$ has at the time $\tau_{0}$ the value $H(\tau_{0})=H(\tau=\tau
_{0})=\,$const. The Hubble function $H$ is usually called \textit{Hubble
constant}.

\textit{Remark}. The Hubble coefficient $H$ has dimension $sec^{-1}$. The
function $H^{-1}$ with dimension $sec$ is usually denoted in astrophysics as
\textit{Hubble's time} \cite{Misner}.

The Hubble function $H$ could also be represented in the forms
\cite{Manoff-7a}
\begin{align}
H  &  =\frac1{n-1}\cdot\theta\mp\sigma(n_{\perp},n_{\perp})=\,\,\nonumber\\
&  =\frac12[(n_{\perp})(h_{u})(\nabla_{u}\overline{g}-\pounds _{u}\overline
{g})(h_{u})(n_{\perp})]\text{ \thinspace\thinspace\thinspace\thinspace.}
\label{1.28p}%
\end{align}

In the Einstein theory of gravitation (ETG) the Hubble coefficient is
considered under the condition that the centrifugal relative velocity is
generated by a shear-free relative velocity in a cosmological model of the
type of Robertson-Walker \cite{Misner}, \cite{Stephani}, i.e.
\begin{equation}
v_{z}=H\cdot l_{\xi_{\perp}}\cdot n_{\perp}=H\cdot\xi_{\perp}\text{
\thinspace\thinspace\thinspace\thinspace\thinspace\thinspace with\thinspace
\thinspace\thinspace\thinspace\thinspace\thinspace\thinspace\thinspace
\thinspace\thinspace\thinspace\thinspace}\sigma=0\text{ \thinspace
\thinspace\thinspace\thinspace.} \label{1.28q}%
\end{equation}

From the explicit form of $H$, it follows that in this case
\[
H=\frac1{n-1}\cdot\theta\,\,\,\,\,
\]
and the Hubble function $H$ depends only on the expansion velocity $\theta$.

\textit{Special case:} $U_{n}$ - or $V_{n}$-space: $\nabla_{u}\overline{g}=0$,
$\pounds _{u}\overline{g}:=0$ (the contravariant vector field $u$ is a Killing
vector field in the corresponding space)
\[
H=0\text{ \thinspace.}
\]

Therefore, in a (pseudo) Riemannian space with or without torsion ($U_{n}$- or
$V_{n}$- space) the Hubble function $H$ is equal to zero if the velocity
vector field $u$ ($n=4$) of an observer is a Killing vector field fulfilling
the Killing equation $\pounds _{u}\overline{g}=0$. This means that the
condition $\pounds _{u}\overline{g}=0$ is a sufficient condition for the
Hubble function $H$ to be equal to zero. Therefore, an observer, moving in
space-time with a velocity vector field $u$ obeying the condition $\pounds
_{\upsilon}g=0$, could not detect any centrifugal (centripetal) relative
velocity $v_{z}$. All mass elements (particles) in the surroundings of the
observer could only circle around him (without any expansion or contraction,
related to the relative centrifugal (centripetal) velocity) and will have
stable orbits with respect to this observer.

\textit{Special case:} $(\overline{L}_{n},g)$-space with $\nabla_{u}%
\overline{g}-\pounds _{u}\overline{g}:=0$ \cite{Manoff-15}. For this case, the
last expression appears as a sufficient condition for the vanishing of the
Hubble function $H$, i.e. $H=0$ if $\nabla_{u}\overline{g}=\pounds _{u}%
\overline{g}$. This is an analogous case for a $(\overline{L}_{n},g)$-space as
the case in $U_{n}$- and $V_{n}$-spaces, where no centrifugal (centripetal)
velocity could be detected.

\textit{Special case}: $(\overline{L}_{n},g)$-spaces with vanishing
centrifugal (centripetal) velocity: $v_{z}:=0$.
\begin{align*}
v_{z}  &  :=0:H=0:\frac1{n-1}\cdot\theta\mp\sigma(n_{\perp},n_{\perp})=0\text{
\thinspace\thinspace\thinspace,}\\
\theta &  =\pm(n-1)\cdot\sigma(n_{\perp},n_{\perp})\text{ \thinspace
\thinspace.}%
\end{align*}

Therefore, if a $(\overline{L}_{n},g)$-space allows the existence of an
observer, who could not detect any centrifugal (centripetal) relative
velocity, the expansion velocity is compensated by the shear velocity in the
space on the basis of the relation $\theta=\pm(n-1)\cdot\sigma(n_{\perp
},n_{\perp})$.

\textit{Remark}. The Hubble function $H$ is introduced in the above
considerations only on a purely kinematic basis related to the notions of
relative velocity and centrifugal (centripetal) relative velocity. Its dynamic
interpretation in a theory of gravitation depends on the structure of the
theory and the relations between the field equations and the Hubble function.

\subsection{Coriolis' velocity}

The vector field
\begin{equation}
v_{c}=\overline{g}[h_{\xi_{\perp}}(_{rel}v)]=g^{ij}\cdot(h_{\xi_{\perp}%
})_{\overline{j}\overline{k}}\cdot~_{rel}v^{k}\cdot\partial_{i}\text{ }
\label{1.28}%
\end{equation}
is called Coriolis' velocity.

\paragraph{Properties of the Coriolis' velocity}

(a) The Coriolis velocity is orthogonal to the vector field $u$, interpreted
as velocity of a mass element (particle), i.e.
\begin{equation}
g(u,v_{c})=0\text{ .} \label{1.29}%
\end{equation}

Proof: From the definition of the Coriolis velocity, it follows
\begin{align}
g(u,v_{c})  &  =g(u,\overline{g}[h_{\xi_{\perp}}(_{rel}v)]=g_{\overline
{i}\overline{j}}\cdot u^{i}\cdot g^{jk}\cdot(h_{\xi_{\perp}})_{\overline
{k}\overline{l}}\cdot~_{rel}v^{l}=\nonumber\\
&  =g_{\overline{i}\overline{j}}\cdot g^{jk}\cdot u^{i}\cdot(h_{\xi_{\perp}%
})_{\overline{k}\overline{l}}\cdot~_{rel}v^{l}=g_{i}^{k}\cdot u^{i}%
\cdot(h_{\xi_{\perp}})_{\overline{k}\overline{l}}\cdot~_{rel}v^{l}%
=\label{1.30}\\
&  =(h_{\xi_{\perp}})_{\overline{k}\overline{l}}\cdot u^{k}\cdot~_{rel}%
v^{l}=(u)(h_{\xi_{\perp}})(_{rel}v)\text{ .}\nonumber
\end{align}

Since
\begin{align}
(u)(h_{\xi_{\perp}})  &  =(h_{\xi_{\perp}})_{\overline{k}\overline{l}}\cdot
u^{k}\cdot dx^{l}=(u)(g)-\frac1{g(\xi_{\perp},\xi_{\perp})}\cdot
(u)[g(\xi_{\perp})]\cdot g(\xi_{\perp})=\nonumber\\
&  =g(u)-\frac1{g(\xi_{\perp},\xi_{\perp})}\cdot g(u,\xi_{\perp})\cdot
g(\xi_{\perp})=g(u)\text{ \ \ , \ }\label{1.31}\\
\text{\ \ }g(u,\xi_{\perp})  &  =0\text{ \ ,\ }\nonumber
\end{align}
then $(u)(h_{\xi_{\perp}})(_{rel}v)=[g(u)](_{rel}v)=g(u,_{rel}v)=0$. Because
of $g(u,_{rel}v)=0$, it follows that $g(u,v_{c})=0$.

(b) The Coriolis velocity $v_{c}$ is orthogonal to the centrifugal
(centripetal) velocity $v_{z}$%
\begin{equation}
g(v_{c},v_{z})=0\text{ .} \label{1.32}%
\end{equation}

(c) The length of the vector $v_{c}$ could be found by the use of the
relations:
\begin{align}
v_{c}  &  =\overline{g}[h_{\xi_{\perp}}(_{rel}v)]~\ \ \ \ \text{,}\nonumber\\
h_{\xi_{\perp}}(_{rel}v)  &  =g(_{rel}v)-\frac1{g(\xi_{\perp},\xi_{\perp}%
)}\cdot[g(\xi_{\perp})](_{rel}v)\cdot g(\xi_{\perp})=\nonumber\\
&  =g(_{rel}v)-\frac{g(\xi_{\perp},_{rel}v)}{g(\xi_{\perp},\xi_{\perp})}\cdot
g(\xi_{\perp})=\label{1.33}\\
&  =g(_{rel}v)\pm g(n_{\perp},\,_{rel}v)\cdot g(n_{\perp})\text{
\ ,}\label{1.33a}\\
\lbrack g(\xi_{\perp})](_{rel}v)  &  =g(\xi_{\perp},_{rel}v)\text{ ,}\nonumber
\end{align}

\begin{align}
v_{c}  &  =\overline{g}[h_{\xi_{\perp}}(_{rel}v)]=~_{rel}v-\frac{g(\xi_{\perp
},_{rel}v)}{g(\xi_{\perp},\xi_{\perp})}\cdot\xi_{\perp}=\label{1.34}\\
&  =_{rel}v\text{ }\pm g(n_{\perp},\,_{rel}v)\cdot n_{\perp}=\,_{rel}%
v-v_{z}\text{ \ \ \ ,} \label{1.34a}%
\end{align}
\begin{align*}
g(v_{c},v_{c})  &  =v_{c}^{2}=\mp l_{v_{c}}^{2}=g(_{rel}v-\frac{g(\xi_{\perp
},_{rel}v)}{g(\xi_{\perp},\xi_{\perp})}\cdot\xi_{\perp},_{rel}v-\frac
{g(\xi_{\perp},_{rel}v)}{g(\xi_{\perp},\xi_{\perp})}\cdot\xi_{\perp})=\\
&  =g(_{rel}v,_{rel}v)-\frac{g(\xi_{\perp},_{rel}v)}{g(\xi_{\perp},\xi_{\perp
})}\cdot g(\xi_{\perp},_{rel}v)-\frac{g(\xi_{\perp},_{rel}v)}{g(\xi_{\perp
},\xi_{\perp})}\cdot g(_{rel}v,\xi_{\perp})+\\
&  +\frac{[g(\xi_{\perp},_{rel}v)]^{2}}{g(\xi_{\perp},\xi_{\perp})}\text{ ,}%
\end{align*}
\begin{equation}
g(v_{c},v_{c})=v_{c}^{2}=\mp l_{v_{c}}^{2}=g(_{rel}v,_{rel}v)-\frac
{[g(\xi_{\perp},_{rel}v)]^{2}}{g(\xi_{\perp},\xi_{\perp})}\text{ \ .}
\label{1.35}%
\end{equation}

On the other side, because of $(u)(h_{u})=h_{u}(u)=0$, and $_{rel}%
v=\overline{g}[h_{u}(\nabla_{u}\xi_{\perp})]$,~~it follows that
\[
g(u,_{rel}v)=h_{u}(u,\nabla_{u}\xi_{\perp})=(u)(h_{u})(\nabla_{u}\xi_{\perp
})\text{ \ ,}
\]

\begin{equation}
g(u,_{rel}v)=(u)(h_{u})(\nabla_{u}\xi_{\perp})=0\text{ .} \label{1.36}%
\end{equation}

Since
\begin{align}
g(_{rel}v,_{rel}v)  &  =~_{rel}v^{2}=\mp l_{_{rel}v}^{2}=g(\overline{g}%
[d(\xi_{\perp})],\overline{g}[d(\xi_{\perp})])=\nonumber\\
&  =\overline{g}(d(\xi_{\perp}),d(\xi_{\perp}))\text{ \ \ \ , \ \ \ \ }%
\pounds _{u}\xi_{\perp}=0\text{ \ ,} \label{1.37}%
\end{align}
\begin{align*}
\overline{g}(d(\xi_{\perp}),d(\xi_{\perp}))  &  =\overline{g}(\sigma
(\xi_{\perp})+\omega(\xi_{\perp})+\frac1{n-1}\cdot\theta\cdot g(\xi_{\perp
}),\\
&  \sigma(\xi_{\perp})+\omega(\xi_{\perp})+\frac1{n-1}\cdot\theta\cdot
g(\xi_{\perp}))
\end{align*}
we obtain%

\begin{align}
g(_{rel}v,_{rel}v)  &  =~_{rel}v^{2}=\overline{g}(d(\xi_{\perp}),d(\xi_{\perp
}))=\nonumber\\
&  =\overline{g}(\sigma(\xi_{\perp}),\sigma(\xi_{\perp}))+\overline{g}%
(\omega(\xi_{\perp}),\omega(\xi_{\perp}))+\nonumber\\
&  +\frac{1}{(n-1)^{2}}\cdot\theta^{2}\cdot g(\xi_{\perp},\xi_{\perp
})+\nonumber\\
&  +2\cdot\overline{g}(\sigma(\xi_{\perp}),\omega(\xi_{\perp}))+\frac{2}%
{n-1}\cdot\theta\cdot\sigma(\xi_{\perp},\xi_{\perp})\text{ .} \label{1.38}%
\end{align}

\textit{Special case:} $\sigma:=0$, $\theta:=0$ (shear-free and expansion-free
velocity).
\begin{equation}
_{rel}v^{2}=\overline{g}(\omega(\xi_{\perp}),\omega(\xi_{\perp}))\text{ .}
\label{1.39}%
\end{equation}

\textit{Special case:} $\omega:=0$ (rotation-free velocity).
\begin{equation}
_{rel}v^{2}=\overline{g}(\sigma(\xi_{\perp}),\sigma(\xi_{\perp}))+\frac
{1}{(n-1)^{2}}\cdot\theta^{2}\cdot g(\xi_{\perp},\xi_{\perp})+\frac{2}%
{n-1}\cdot\theta\cdot\sigma(\xi_{\perp},\xi_{\perp})\text{ .} \label{1.40}%
\end{equation}

\textit{Special case:} $\theta:=0$, $\omega:=0$ (expansion-free and
rotation-free velocity).
\begin{equation}
_{rel}v^{2}=\overline{g}(\sigma(\xi_{\perp}),\sigma(\xi_{\perp}))\text{ \ .}
\label{1.41}%
\end{equation}

\textit{Special case:} $\sigma:=0$, $\omega:=0$ (shear-free and rotation-free
velocity).
\begin{equation}
_{rel}v^{2}=\frac{1}{(n-1)^{2}}\cdot\theta^{2}\cdot g(\xi_{\perp},\xi_{\perp
})\text{ \ .} \label{1.42}%
\end{equation}

\textit{Special case:} $\theta:=0$ (expansion-free velocity).
\begin{equation}
_{rel}v^{2}=\overline{g}(\sigma(\xi_{\perp}),\sigma(\xi_{\perp}))+\overline
{g}(\omega(\xi_{\perp}),\omega(\xi_{\perp}))+2\cdot\overline{g}(\sigma
(\xi_{\perp}),\omega(\xi_{\perp}))\text{ .} \label{1.43}%
\end{equation}

The square $v_{c}^{2}$ of $v_{c}$ could be found on the basis of the relation
$v_{c}^{2}=~_{rel}v^{2}-v_{z}^{2}$%
\begin{align}
v_{c}^{2}  &  =~_{rel}v^{2}-v_{z}^{2}=\overline{g}(\sigma(\xi_{\perp}%
),\sigma(\xi_{\perp}))-\frac{[\sigma(\xi_{\perp},\xi_{\perp})]^{2}}%
{g(\xi_{\perp},\xi_{\perp})}+\nonumber\\
&  +\overline{g}(\omega(\xi_{\perp}),\omega(\xi_{\perp}))+2\cdot\overline
{g}(\sigma(\xi_{\perp}),\omega(\xi_{\perp}))\text{ .} \label{1.44}%
\end{align}

Therefore, the length of $v_{c}$ does not depend on the expansion velocity
invariant $\theta$.

\textit{Special case:} $\sigma:=0$ (shear-free velocity).
\begin{equation}
v_{c}^{2}=\overline{g}(\omega(\xi_{\perp}),\omega(\xi_{\perp}))\text{ .}
\label{1.45}%
\end{equation}

\textit{Special case:} $\omega:=0$ (rotation-free velocity).
\begin{equation}
v_{c}^{2}=\overline{g}(\sigma(\xi_{\perp}),\sigma(\xi_{\perp}))-\frac
{[\sigma(\xi_{\perp},\xi_{\perp})]^{2}}{g(\xi_{\perp},\xi_{\perp})}\text{ .}
\label{1.46}%
\end{equation}

It follows from the last (previous) expression that even if the rotation
velocity tensor $\omega$ is equal to zero $(\omega=0)$, the Coriolis
(relative) velocity $v_{c}$ is not equal to zero if $\sigma\neq0$.

(d) The scalar product between $v_{c}$ and $_{rel}v$ could be found in its
explicit form by the use of the relations:
\begin{equation}
g(v_{c},_{rel}v)=h_{\xi_{\perp}}(_{rel}v,_{rel}v)=~_{rel}v^{2}-\frac
{[g(\xi_{\perp},_{rel}v)]^{2}}{g(\xi_{\perp},\xi_{\perp})}=~_{rel}v^{2}%
-v_{z}^{2}=v_{c}^{2}\text{ .} \label{1.47}%
\end{equation}

(e) The explicit form of $v_{c}$ could be found by the use of the relations
\[
v_{c}=\overline{g}[h_{\xi_{\perp}}(_{rel}v)]=~_{rel}v-\frac{g(\xi_{\perp
},_{rel}v)}{g(\xi_{\perp},\xi_{\perp})}\cdot\xi_{\perp}=~_{rel}v-v_{z}\text{
,}
\]
\begin{equation}
v_{c}=\overline{g}[\sigma(\xi_{\perp})]-\frac{\sigma(\xi_{\perp},\xi_{\perp}%
)}{g(\xi_{\perp},\xi_{\perp})}\cdot\xi_{\perp}+\overline{g}[\omega(\xi_{\perp
})]\text{ .} \label{1.48}%
\end{equation}

Therefore, the Coriolis velocity does not depend on the expansion velocity
invariant $\theta$.

\textit{Special case:} $\sigma:=0$ (shear-free velocity).
\begin{equation}
v_{c}=\overline{g}[\omega(\xi_{\perp})]\text{ .} \label{1.49}%
\end{equation}

\textit{Special case:} $\omega:=0$ (rotation-free velocity).
\begin{equation}
v_{c}=\overline{g}[\sigma(\xi_{\perp})]-\frac{\sigma(\xi_{\perp},\xi_{\perp}%
)}{g(\xi_{\perp},\xi_{\perp})}\cdot\xi_{\perp}\text{ .} \label{1.50}%
\end{equation}

(f) The Coriolis velocity $v_{c}$ is orthogonal to the deviation vector
$\xi_{\perp}$, i.e. $g(v_{c},\xi_{\perp})=0$.

Proof: From $g(v_{c},\xi_{\perp})=g(\overline{g}[h_{\xi_{\perp}}(_{rel}%
v)],\xi_{\perp})=h_{\xi_{\perp}}(\xi_{\perp},_{rel}v)=(\xi_{\perp}%
)(h_{\xi_{\perp}})(_{rel}v)$ and $(\xi_{\perp})(h_{\xi_{\perp}})=0$, it
follows that
\begin{equation}
g(v_{c},\xi_{\perp})=0. \label{1.51}%
\end{equation}

\section{Centrifugal (centripetal) and Coriolis' accelerations}

In analogous way as in the case of centrifugal (centripetal) and Coriolis'
velocities, the corresponding accelerations could be defined by the use of the
projections of the relative acceleration $_{rel}a=\overline{g}[h_{u}%
(\nabla_{u}\nabla_{u}\xi_{\perp})]$ along or orthogonal to the vector field
$\xi_{\perp}$%
\begin{equation}
_{rel}a=\frac{g(\xi_{\perp},~_{rel}a)}{g(\xi_{\perp},\xi_{\perp})}\cdot
\xi_{\perp}+\overline{g}[h_{\xi_{\perp}}(_{rel}a)]=a_{z}+a_{c}\text{ \ ,}
\label{2.1}%
\end{equation}
where
\begin{equation}
a_{z}=\frac{g(\xi_{\perp},~_{rel}a)}{g(\xi_{\perp},\xi_{\perp})}\cdot
\xi_{\perp}\text{ \ ,} \label{2.2}%
\end{equation}
\begin{equation}
a_{c}=\overline{g}[h_{\xi_{\perp}}(_{rel}a)]\text{ \ .} \label{2.3}%
\end{equation}

If
\begin{equation}
\frac{g(\xi_{\perp},~_{rel}a)}{g(\xi_{\perp},\xi_{\perp})}>0 \label{2.4}%
\end{equation}
the vector field $a_{z}$ is called (relative) \textit{centrifugal
acceleration.} If
\begin{equation}
\frac{g(\xi_{\perp},~_{rel}a)}{g(\xi_{\perp},\xi_{\perp})}<0 \label{2.5}%
\end{equation}
the vector field $a_{z}$ is called (relative)\textit{\ centripetal
acceleration.}

The vector field $a_{c}$ is called (relative) \textit{Coriolis' acceleration.}

\subsection{Centrifugal (centripetal) acceleration}

The centrifugal (centripetal) acceleration has specific properties which will
be considered below on the basis of the properties of the relative
acceleration $_{rel}a$.

The relative acceleration $_{rel}a$ is orthogonal to the vector field $u.$

Proof: From $g(u,_{rel}a)=g(u,\overline{g}[h_{u}(\nabla_{u}\nabla_{u}%
\xi_{\perp})])=(u)(h_{u})(\nabla_{u}\nabla_{u}\xi_{\perp})$ and $(u)(h_{u}%
)=h_{u}(u)=0$, it follows that
\begin{equation}
g(u,_{rel}a)=0\text{ .} \label{2.6}%
\end{equation}

(a) The centrifugal (centripetal) acceleration $a_{z}$ is orthogonal to the
vector field $u$, i.e.
\begin{equation}
g(u,a_{z})=0. \label{2.6a}%
\end{equation}

Proof: From
\begin{align}
g(u,a_{z})  &  =g(u,\frac{g(\xi_{\perp},~_{rel}a)}{g(\xi_{\perp},\xi_{\perp}%
)}\cdot\xi_{\perp})=\nonumber\\
&  =\frac{g(\xi_{\perp},~_{rel}a)}{g(\xi_{\perp},\xi_{\perp})}\cdot
g(u,\xi_{\perp})\text{ , ~}\label{2.7}\\
\text{\ \ \ \ }g(u,\xi_{\perp})  &  =0\text{ ,}\nonumber
\end{align}
it follows that $g(u,a_{z})=0$.

(b) The centrifugal (centripetal) acceleration $a_{z}$ is orthogonal to the
Coriolis acceleration $a_{c}$, i.e.
\begin{equation}
g(a_{z},a_{c})=0\text{ .} \label{2.8}%
\end{equation}

Proof: From
\[
g(\frac{g(\xi_{\perp},~_{rel}a)}{g(\xi_{\perp},\xi_{\perp})}\cdot\xi_{\perp
},\overline{g}[h_{\xi_{\perp}}(_{rel}a)])=\frac{g(\xi_{\perp},~_{rel}a)}%
{g(\xi_{\perp},\xi_{\perp})}\cdot g(\xi_{\perp},\overline{g}[h_{\xi_{\perp}%
}(_{rel}a)])\text{ ,}
\]
\begin{align}
g(\xi_{\perp},\overline{g}[h_{\xi_{\perp}}(_{rel}a)])  &  =(\xi_{\perp
})(h_{\xi_{\perp}})(_{rel}a)\text{ ,}\label{2.9}\\
(\xi_{\perp})(h_{\xi_{\perp}})  &  =(h_{\xi_{\perp}})(\xi_{\perp})=0\text{ ,}
\label{2.10}%
\end{align}
it follows that $g(a_{z},a_{c})=0$.

(c) The length of the vector $a_{z}$ could be found on the basis of the
relations
\begin{equation}
a_{z}^{2}=g(a_{z},a_{z})=\mp l_{a_{z}}^{2}=\frac{[g(\xi_{\perp},~_{rel}%
a)]^{2}}{g(\xi_{\perp},\xi_{\perp})}\text{ \ \ \ .} \label{2.11}%
\end{equation}

Therefore, in general, the square $a_{z}^{2}$ of the length of the centrifugal
(centripetal) acceleration $a_{z}$ is reverse proportional to $\xi_{\perp}%
^{2}=g(\xi_{\perp},\xi_{\perp})$.

\textit{Special case:} $M_{n}=E_{n}$, $n=3$ (3-dimensional Euclidean space).
\begin{align}
\xi_{\perp}  &  :=\overrightarrow{r}\text{ , \ \ \ }g(\xi_{\perp},\xi_{\perp
})=r^{2}\text{ \ , \ \ \ \ }_{rel}a=~_{rel}\overrightarrow{a}\text{\ }%
,\nonumber\\
a_{z}^{2}  &  =\frac{[g(\overrightarrow{r},~_{rel}\overrightarrow{a})]^{2}%
}{r^{2}}\text{ , \ \ \ \ \ }l_{a_{z}}=\frac{g(\overrightarrow{r}%
,\,_{rel}\overrightarrow{a})}r=g(n_{\perp},\,_{rel}\overrightarrow{a})\text{
,}\label{2.12}\\
n_{\perp}  &  :=\frac{\overrightarrow{r}}r\text{ , \ \ \ \ \ \ }g(n_{\perp
},n_{\perp})=\frac{g(\overrightarrow{r},\overrightarrow{r})}{r^{2}}=n_{\perp
}^{2}=1\text{ \ \ .} \label{2.12a}%
\end{align}

The length $l_{a_{z}}=\mid g(a_{z},a_{z})\mid^{1/2}$ of the centrifugal
(centripetal) acceleration $a_{z}$ is equal to the projection of the relative
acceleration $_{rel}\overrightarrow{a}$ at the unit vector field $n_{\perp}$
along the vector field $\xi_{\perp}$. If \ $l_{a_{z}}=g(n_{\perp}%
,\,_{rel}\overrightarrow{a})=0$, i.e. if the relative acceleration
$_{rel}\overrightarrow{a}$ is orthogonal to the radius vector $\overrightarrow
{r}$ [$_{rel}\overrightarrow{a}\perp\overrightarrow{r}$], then $a_{z}=0$.

If the relative acceleration $_{rel}a$ is equal to zero then the centrifugal
(centripetal) acceleration $a_{z}$ is also equal to zero.

(d) The scalar product $g(\xi_{\perp},_{rel}a)$ could be found in its explicit
form by the use of the explicit form of $_{rel}a$ \cite{Manoff-5}
\begin{equation}
_{rel}a=\overline{g}[A(\xi_{\perp})]=\overline{g}[_{s}D(\xi_{\perp
})]+\overline{g}[W(\xi_{\perp})]+\frac1{n-1}\cdot U\cdot\xi_{\perp}\text{ .}
\label{2.13}%
\end{equation}

Then
\begin{align}
g(\xi_{\perp},~_{rel}a)  &  =~_{s}D(\xi_{\perp},\xi_{\perp})+\frac{1}%
{n-1}\cdot U\cdot g(\xi_{\perp},\xi_{\perp})\text{ ,}\label{2.14}\\
a_{z}^{2}  &  =\frac{[g(\xi_{\perp},~_{rel}a)]^{2}}{g(\xi_{\perp},\xi_{\perp
})}=\nonumber\\
&  =\frac{1}{g(\xi_{\perp},\xi_{\perp})}\cdot\lbrack_{s}D(\xi_{\perp}%
,\xi_{\perp})+\frac{1}{n-1}\cdot U\cdot g(\xi_{\perp},\xi_{\perp}%
)]^{2}=\nonumber\\
&  =\frac{[_{s}D(\xi_{\perp},\xi_{\perp})]^{2}}{g(\xi_{\perp},\xi_{\perp}%
)}+\frac{1}{(n-1)^{2}}\cdot U^{2}\cdot g(\xi_{\perp},\xi_{\perp})+\nonumber\\
&  +\frac{2}{n-1}\cdot U\cdot~_{s}D(\xi_{\perp},\xi_{\perp})\text{ .}
\label{2.15}%
\end{align}

\textit{Special case:} $_{s}D:=0$ (shear-free acceleration).
\begin{equation}
a_{z}^{2}=\frac{1}{(n-1)^{2}}\cdot U^{2}\cdot g(\xi_{\perp},\xi_{\perp})\text{
\ .} \label{2.16}%
\end{equation}

\textit{Special case:} $U:=0$ (expansion-free acceleration).
\begin{equation}
a_{z}^{2}=\frac{[_{s}D(\xi_{\perp},\xi_{\perp})]^{2}}{g(\xi_{\perp},\xi
_{\perp})}\text{ \ .} \label{2.17}%
\end{equation}

(e) The explicit form of $a_{z}$ could be found in the form
\begin{align}
a_{z}  &  =\frac{g(\xi_{\perp},~_{rel}a)}{g(\xi_{\perp},\xi_{\perp})}\cdot
\xi_{\perp}=\frac{_{s}D(\xi_{\perp},\xi_{\perp})}{g(\xi_{\perp},\xi_{\perp}%
)}\cdot\xi_{\perp}+\frac1{n-1}\cdot U\cdot\xi_{\perp}=\nonumber\\
&  =[\frac1{n-1}\cdot U+\frac{_{s}D(\xi_{\perp},\xi_{\perp})}{g(\xi_{\perp
},\xi_{\perp})}]\cdot\xi_{\perp}=\label{2.18}\\
&  =[\frac1{n-1}\cdot U\pm\,_{s}D(n_{\perp},n_{\perp})]\cdot\xi_{\perp}\text{
\ \ \ .} \label{2.18a}%
\end{align}

If
\begin{equation}
\frac1{n-1}\cdot U+\frac{_{s}D(\xi_{\perp},\xi_{\perp})}{g(\xi_{\perp}%
,\xi_{\perp})}>0\text{ } \label{2.19}%
\end{equation}
$a_{z}$ is a centrifugal (relative) acceleration. If
\begin{equation}
\frac1{n-1}\cdot U+\frac{_{s}D(\xi_{\perp},\xi_{\perp})}{g(\xi_{\perp}%
,\xi_{\perp})}<0 \label{2.20}%
\end{equation}
$a_{z}$ is a centripetal (relative) acceleration.

The (relative) centrifugal (centripetal) acceleration $a_{z}$ could be written
also in the form
\begin{equation}
a_{z}=\overline{q}\cdot\xi_{\perp}\,\,\,\,\,\text{,} \label{2.21}%
\end{equation}
where
\begin{equation}
\overline{q}=\frac1{n-1}\cdot U\pm\,_{s}D(n_{\perp},n_{\perp})\,\,\,\text{.}
\label{2.21a}%
\end{equation}

The invariant function $\overline{q}$ could be denoted as \textit{acceleration
function} or \textit{acceleration parameter}. Its explicit form depends on the
explicit forms of the expansion acceleration invariant $U$ and on the shear
acceleration tensor $_{s}D$ \cite{Manoff-5}, \cite{Manoff-7}. The centrifugal
(centripetal) acceleration $a_{z}$ has an analogous form to the centrifugal
(centripetal) velocity $v_{z}$%
\begin{align}
a_{z}  &  =\overline{q}\cdot l_{\xi_{\perp}}\cdot n_{\perp}=\pm l_{a_{z}}\cdot
n_{\perp}\text{ \thinspace\thinspace,}\label{2.21b}\\
v_{z}  &  =H\cdot l_{\xi_{\perp}}\cdot n_{\perp}=\pm l_{v_{z}}\cdot n_{\perp
}\,\,\,\,\text{.} \label{2.21c}%
\end{align}

Since $\pm l_{v_{z}}=H\cdot l_{\xi_{\perp}}$ and $\pm l_{a_{z}}=\overline
{q}\cdot l_{\xi_{\perp}}$, we can express $l_{\xi_{\perp}}$ from the one of
these expressions
\begin{equation}
l_{\xi_{\perp}}=\pm\frac{l_{v_{z}}}H=\pm\frac{l_{a_{z}}}{\overline{q}}\text{ }
\label{2.21d}%
\end{equation}
and put it into the other relation. As a result we could find the relation
between the absolute values of $a_{z}$ and $v_{z}$%
\begin{equation}
l_{a_{z}}=\frac{\overline{q}}H\cdot l_{v_{z}}\text{ .} \label{2.21e}%
\end{equation}

Therefore, the absolute value $l_{a_{z}}$ of the centrifugal (centripetal)
acceleration $a_{z}$ could be related to the absolute value $l_{v_{z}}$ of the
centrifugal (centripetal) velocity $v_{z}$ by means of the Hubble's function
$H $ and the acceleration parameter $\overline{q}$. Then
\begin{align}
a_{z}  &  =\pm\frac{\overline{q}}H\cdot l_{v_{z}}\cdot n_{\perp}%
=\frac{\overline{q}}H\cdot v_{z}\,\,\,\,\,\text{,}\label{2.21f}\\
\overline{q}  &  =H\cdot\frac{l_{a_{z}}}{l_{v_{z}}}=\pm\frac{l_{a_{z}}}%
{l_{\xi_{\perp}}}\,\,\,\,\,\,\text{.} \label{2.21g}%
\end{align}

\textit{Remark}. In Einstein's theory of gravitation, on the basis of
cosmological models with Robertson-Walker metrics, the acceleration parameter
is introduced as \cite{Stephani}
\[
q\sim\frac1H\cdot\frac{l_{a_{z}}}{l_{v_{z}}}\,\,\text{,}
\]
where $l_{a_{z}}=\ddot{K}$, $H=\dot{K}/K$, $l_{v_{z}}=\dot{K}$, $l_{\xi
_{\perp}}=K$. A comparison with the expression for $\overline{q}$ shows that
$\overline{q}\sim q\cdot H^{2}$, and $q\sim\overline{q}/H^{2}$. It should be
stressed that the acceleration parameter $\overline{q}$ is introduced on a
pure kinematic basis and is not depending on a special type of gravitational
theory in a $(\overline{L}_{n},g)$-space (Einstein's theory of gravitation
could be considered as a gravitational theory in a special type of a
$(\overline{L}_{n},g)$-space with $n=4$ and $(\overline{L}_{n},g)\equiv V_{4}$).

In a theory of gravitation the centripetal (relative) acceleration could be
interpreted as gravitational acceleration.

\textit{Special case:} $_{s}D:=0$ (shear-free relative acceleration).
\begin{equation}
a_{z}=\frac{1}{n-1}\cdot U\text{ }\cdot\xi_{\perp}\text{ \ \ \ \ .}
\label{2.22}%
\end{equation}

If the expansion acceleration invariant $U>0$ the acceleration $a_{z}$ is a
centrifugal (or expansion) acceleration. If $U<0$ the acceleration $a_{z}$ is
a centripetal (or contraction) acceleration. Therefore, in the case of a
shear-free relative acceleration the centrifugal or the centripetal
acceleration is proportional to the expansion acceleration invariant $U$.

\subsection{Coriolis' acceleration}

The vector field $a_{c}$, defined as
\[
a_{c}=\overline{g}[h_{\xi_{\perp}}(_{rel}a)]\text{ ,}
\]
is called Coriolis' (relative) acceleration. On the basis of its definition,
the Coriolis acceleration has well defined properties.

(a) The Coriolis acceleration is orthogonal to the vector field $u$, i.e.
\begin{equation}
g(u,a_{c})=0. \label{2.23}%
\end{equation}

Proof: From
\[
g(u,a_{c})=g(u,\overline{g}[h_{\xi_{\perp}}(_{rel}a)])=(u)(h_{\xi_{\perp}%
})(_{rel}a)=h_{\xi_{\perp}}(u,_{rel}a)
\]
and
\begin{equation}
(u)(h_{\xi_{\perp}})=(h_{\xi_{\perp}})(u)=g(u)\text{ \ , \ \ \ \ }%
g(u,_{rel}a)=0\text{ \ ,} \label{2.23a}%
\end{equation}
it follows that
\[
g(u,a_{c})=[g(u)](_{rel}a)=g(u,_{rel}a)=0.
\]

(b) The Coriolis acceleration $a_{c}$ is orthogonal to the centrifugal
(centripetal) acceleration $a_{z}$, i.e.
\begin{equation}
g(a_{c},a_{z})=0\text{ \ .} \label{2.24}%
\end{equation}

(c) The Coriolis acceleration $a_{c}$ is orthogonal to the deviation vector
$\xi_{\perp}$, i.e.
\begin{equation}
g(\xi_{\perp},a_{c})=0\text{ \ \ .} \label{2.25}%
\end{equation}

(d) The length $\sqrt{\mid a_{c}^{2}\mid}=\sqrt{\mid g(a_{c},a_{c})\mid}$ of
$a_{c}$ could be found by the use of the relations
\begin{align}
a_{c}  &  =\overline{g}[h_{\xi_{\perp}}(_{rel}a)]\text{ ,}\nonumber\\
\lbrack g(\xi_{\perp})](_{rel}a)  &  =g(\xi_{\perp},_{rel}a)\text{
\ ,}\label{2.26}\\
h_{\xi_{\perp}}(_{rel}a)  &  =g(_{rel}a)-\frac{g(\xi_{\perp},_{rel}a)}%
{g(\xi_{\perp},\xi_{\perp})}\cdot g(\xi_{\perp})\text{ \ ,} \label{2.27}%
\end{align}

\begin{align}
a_{c}  &  =\overline{g}[h_{\xi_{\perp}}(_{rel}a)]=~_{rel}a-a_{z}%
=\label{2.28}\\
&  =~_{rel}a-\frac{g(\xi_{\perp},_{rel}a)}{g(\xi_{\perp},\xi_{\perp})}\cdot
\xi_{\perp}\text{ \ \ \ . } \label{2.29}%
\end{align}

For $a_{c}^{2}$ we obtain
\begin{align}
a_{c}^{2}  &  =g(a_{c},a_{c})=g(_{rel}a-a_{z},~_{rel}a-a_{z})=\nonumber\\
&  =g(_{rel}a,_{rel}a)+g(a_{z},a_{z})-2\cdot g(a_{z},_{rel}a)\text{ \ .}
\label{2.30}%
\end{align}

Since
\begin{equation}
g(a_{z},a_{z})=g(a_{z},_{rel}a)=\frac{[g(\xi_{\perp}),_{rel}a)]^{2}}%
{g(\xi_{\perp},\xi_{\perp})\text{ }}\text{ \ \ ,} \label{2.31}%
\end{equation}
\begin{equation}
a_{c}^{2}=g(_{rel}a,_{rel}a)-g(a_{z},a_{z})=~_{rel}a^{2}-a_{z}^{2}\text{
\ \ ,} \label{2.32}%
\end{equation}
\begin{align*}
g(_{rel}a,_{rel}a)  &  =~_{rel}a^{2}=g(\overline{g}[A(\xi_{\perp}%
)],\overline{g}[A(\xi_{\perp})])=\\
&  =\overline{g}(A(\xi_{\perp}),A(\xi_{\perp}))\text{ ,}\\
A  &  =~_{s}D+W+\frac{1}{n-1}\cdot U\cdot h_{u}\text{ \ \ ,}\\
A(\xi_{\perp})  &  =~_{s}D(\xi_{\perp})+W(\xi_{\perp})+\frac{1}{n-1}\cdot
U\cdot g(\xi_{\perp})\text{ \ ,}\\
h_{u}(\xi_{\perp})  &  =g(\xi_{\perp})\text{ \ ,}%
\end{align*}

\begin{align}
\overline{g}(_{s}D(\xi_{\perp}),h_{u}(\xi_{\perp}))  &  =~_{s}D(\xi_{\perp
},\xi_{\perp})\text{ \ ,}\label{2.33}\\
\overline{g}(W(\xi_{\perp}),h_{u}(\xi_{\perp}))  &  =~W(\xi_{\perp},\xi
_{\perp})=0\text{ \ \ ,}\label{2.34}\\
\overline{g}(h_{u}(\xi_{\perp}),h_{u}(\xi_{\perp}))  &  =~h_{u}(\xi_{\perp
},\xi_{\perp})=g\text{ }(\xi_{\perp},\xi_{\perp})\text{ , } \label{2.35}%
\end{align}
it follows for $_{rel}a^{2}$%
\begin{align}
_{rel}a^{2}  &  =g(_{rel}a,_{rel}a)=\overline{g}(_{s}D(\xi_{\perp}),_{s}%
D(\xi_{\perp}))+\overline{g}(W(\xi_{\perp}),W(\xi_{\perp}))+\nonumber\\
&  +2\cdot\overline{g}(_{s}D(\xi_{\perp}),W(\xi_{\perp}))+\nonumber\\
&  +\frac{2}{n-1}\cdot U\cdot~_{s}D(\xi_{\perp},\xi_{\perp})+\frac
{1}{(n-1)^{2}}\cdot U^{2}\cdot g\text{ }(\xi_{\perp},\xi_{\perp})\text{ .}
\label{2.36}%
\end{align}

On the other side,
\[
a_{z}^{2}=\frac{[_{s}D(\xi_{\perp},\xi_{\perp})]^{2}}{g(\xi_{\perp},\xi
_{\perp})}+\frac{1}{(n-1)^{2}}\cdot U^{2}\cdot g(\xi_{\perp},\xi_{\perp
})+\frac{2}{n-1}\cdot U\cdot~_{s}D(\xi_{\perp},\xi_{\perp})\text{ .}
\]

Therefore,
\begin{align}
a_{c}^{2}  &  =~_{rel}a^{2}-a_{z}^{2}=\nonumber\\
&  =\overline{g}(_{s}D(\xi_{\perp}),_{s}D(\xi_{\perp}))-\frac{[_{s}%
D(\xi_{\perp},\xi_{\perp})]^{2}}{g(\xi_{\perp},\xi_{\perp})}+\nonumber\\
&  +\overline{g}(W(\xi_{\perp}),W(\xi_{\perp}))+2\cdot\overline{g}(_{s}%
D(\xi_{\perp}),W(\xi_{\perp}))\text{ \ .} \label{2.37}%
\end{align}

\textit{Special case:} $_{s}D:=0$ (shear-free acceleration).
\begin{equation}
a_{c}^{2}=\overline{g}(W(\xi_{\perp}),W(\xi_{\perp}))\text{ \ \ .}
\label{2.38}%
\end{equation}

\textit{Special case:} $W:=0$ (rotation-free acceleration).
\begin{equation}
a_{c}^{2}=\overline{g}(_{s}D(\xi_{\perp}),_{s}D(\xi_{\perp}))-\frac{[_{s}%
D(\xi_{\perp},\xi_{\perp})]^{2}}{g(\xi_{\perp},\xi_{\perp})}\text{ \ \ .}
\label{2.39}%
\end{equation}

The explicit form of $a_{c}$ could be found by the use of the relations:
\begin{align}
a_{c}  &  =\overline{g}[h_{\xi_{\perp}}(_{rel}a)]=~_{rel}a-\frac{g(\xi_{\perp
},_{rel}a)}{g(\xi_{\perp},\xi_{\perp})}\cdot\xi_{\perp}=~_{rel}a-a_{z}%
=\nonumber\\
&  =\overline{g}[_{s}D(\xi_{\perp})]-\frac{_{s}D(\xi_{\perp},\xi_{\perp}%
)}{g(\xi_{\perp},\xi_{\perp})}\cdot\xi_{\perp}+\overline{g}[W(\xi_{\perp
})]=\label{2.40}\\
&  =\overline{g}[_{s}D(\xi_{\perp})]\mp\,_{s}D(n_{\perp},n_{\perp})\cdot
\xi_{\perp}+\overline{g}[W(\xi_{\perp})]\text{ .} \label{2.40a}%
\end{align}

Therefore, the Coriolis acceleration $a_{c}$ does not depend on the expansion
acceleration invariant $U$.

\textit{Special case:} $_{s}D:=0$ (shear-free acceleration).
\begin{equation}
a_{c}=\overline{g}[W(\xi_{\perp})]\text{ \ \ .} \label{2.41}%
\end{equation}

\textit{Special case:} $W:=0$ (rotation-free acceleration).
\begin{align}
a_{c}  &  =\overline{g}[_{s}D(\xi_{\perp})]-\frac{_{s}D(\xi_{\perp},\xi
_{\perp})}{g(\xi_{\perp},\xi_{\perp})}\cdot\xi_{\perp}=\label{2.42}\\
&  =\overline{g}[_{s}D(\xi_{\perp})]\mp\,_{s}D(n_{\perp},n_{\perp})\cdot
\xi_{\perp}\text{ \ .} \label{2.42a}%
\end{align}

The Coriolis acceleration depends on the shear acceleration $_{s}D$ and on the
rotation acceleration $W$. These types of accelerations generate a Coriolis
acceleration between particles or mass elements in a flow.

\section{Centrifugal (centripetal) acceleration as gravitational acceleration}

1. The main idea of the Einstein theory of gravitation (ETG) is the
identification of the centripetal acceleration with the gravitational
acceleration. The weak equivalence principle stays that a gravitational
acceleration could be identified with a centripetal acceleration and vice
versa. From this point of view, it is worth to be investigated the relation
between the centrifugal (centripetal) acceleration and the Einstein theory of
gravitation as well as the possibility for describing the gravitational
interaction as result of the centrifugal (centripetal) acceleration generated
by the motion of mass elements (particles).

The structure of the centrifugal (centripetal) acceleration could be
considered on the basis of its explicit form expressed by means of the
kinematic characteristics of the relative acceleration and the relative
velocity. The centrifugal (centripetal) acceleration is written as
\[
a_{z}=[\frac{1}{n-1}\cdot U+\frac{_{s}D(\xi_{\perp},\xi_{\perp})}{g(\xi
_{\perp},\xi_{\perp})}]\cdot\xi_{\perp}\text{ .}
\]

The vector field $\xi_{\perp}$ is directed outside of the trajectory of a mass
element (particle). If the mass element generates a gravitational field the
acceleration $a_{z}$ should be in the direction to the mass element. If the
mass element moves in an external gravitational field caused by an other
gravitational source then the acceleration $a_{z}$ should be directed to the
source. The expansion (contraction) invariant $U$ could be expressed by means
of the kinematic characteristics of the relative acceleration or of the
relative velocity in the forms \cite{Manoff-3}, \cite{Manoff-5}
\begin{equation}
U=U_{0}+I=~_{F}U_{0}-~_{T}U_{0}+I\text{ \ , \ \ \ \ \ \ \ \ }U_{0}=~_{F}%
U_{0}-~_{T}U_{0}\text{ ,}~ \label{3.1}%
\end{equation}
where $_{F}U_{0}$ is the torsion-free and curvature-free expansion
acceleration, $_{T}U_{0}$ is the expansion acceleration induced by the torsion
and $I$ is the expansion acceleration induced by the curvature, $U_{0}$ is the
curvature-free expansion acceleration
\begin{align}
_{F}U_{0}  &  =g[b]-\frac{1}{e}\cdot g(u,\nabla_{u}a)\text{ ,}\label{3.2}\\
_{F}U_{0}  &  =a^{k}~_{;k}-\frac{1}{e}\cdot g_{\overline{k}\overline{l}}\cdot
u^{k}\cdot a^{l}~_{;m}\cdot u^{m}\text{ ,} \label{3.3}%
\end{align}
\begin{align}
U_{0}  &  =g[b]-\overline{g}[_{s}P(\overline{g})\sigma]-\overline
{g}[Q(\overline{g})\omega]-\dot{\theta}_{1}-\frac{1}{n-1}\cdot\theta_{1}%
\cdot\theta-\nonumber\\
&  -\frac{1}{e}\cdot\lbrack g(u,T(a,u))+g(u,\nabla_{u}a)]\text{ ,} \label{3.4}%
\end{align}
\begin{equation}
I=R_{ij}\cdot u^{i}\cdot u^{j}\text{ \ , \ \ \ \ \ }g[b]=g_{\overline
{i}\overline{j}}\cdot b^{ij}=g_{\overline{i}\overline{j}}\cdot a^{i}%
~_{;n}\cdot g^{nj}=a^{n}~_{;n}\text{ .} \label{3.5}%
\end{equation}

2. In the ETG only the term $I$ is used on the basis of the Einstein
equations. The invariant $I$ represents an invariant generalization of
Newton's gravitational law \cite{Manoff-8}. In a $V_{n}$-space $(n=4)$ of the
ETG, a free moving spinless test particle with $a=0$ will have an expansion
(contraction) acceleration $U=I$ $(U_{0}=0)$ if $R_{ij}\neq0$ and $U=I=0$ if
$R_{ij}=0$. At the same time, in a $V_{n}$-space (the bars over the indices
should be omitted)%

\begin{align}
_{s}D  &  =~_{s}M\text{ , \ \ \ \ \ }_{s}D_{0}=0\text{ , \ \ \ \ }%
M=h_{u}(K_{s})h_{u}\text{ ,}\label{3.6}\\
M_{ij}  &  =h_{i\overline{k}}\cdot K_{s}^{kl}\cdot h_{\overline{l}j}=\frac
{1}{2}\cdot h_{i\overline{k}}\cdot(K_{s}^{kl}+K_{s}^{lk})\cdot h_{\overline
{l}j}=\nonumber\\
&  =\frac{1}{2}\cdot h_{i\overline{k}}\cdot(R^{k}~_{mnr}\cdot u^{m}\cdot
u^{n}\cdot g^{rl}+R^{l}~_{mnr}\cdot u^{m}\cdot u^{n}\cdot g^{rk})\cdot
h_{\overline{l}j}=\nonumber\\
&  =\frac{1}{2}\cdot u^{m}\cdot u^{n}\cdot(h_{i\overline{k}}\cdot R^{k}%
~_{mnr}\cdot g^{rl}\cdot h_{\overline{l}j}+h_{i\overline{k}}\cdot R^{l}%
~_{mnr}\cdot g^{rk}\cdot h_{\overline{l}j})=\nonumber\\
&  =\frac{1}{2}\cdot u^{m}\cdot u^{n}\cdot\lbrack h_{i\overline{k}}\cdot
R^{k}~_{mnr}\cdot g^{rl}\cdot(g_{\overline{l}j}-\frac{1}{e}\cdot
u_{\overline{l}}\cdot u_{j})+\nonumber\\
&  +h_{\overline{l}j}\cdot R^{k}~_{mnr}\cdot g^{rk}\cdot(g_{\overline{k}%
i}-\frac{1}{e}\cdot u_{\overline{k}}\cdot u_{i})\text{ \ .} \label{3.7}%
\end{align}

Since
\[
h_{i\overline{k}}\cdot R^{k}~_{mnr}\cdot g^{rl}\cdot(g_{\overline{l}j}%
-\frac1e\cdot u_{\overline{l}}\cdot u_{j})=
\]
\begin{align*}
&  =(g_{i\overline{k}}-\frac1e\cdot u_{i}\cdot u_{\overline{k}})\cdot
R^{k}~_{mnr}\cdot g^{rl}\cdot g_{\overline{l}j}-\\
&  -\frac1e\cdot(g_{i\overline{k}}-\frac1e\cdot u_{i}\cdot u_{\overline{k}%
})\cdot R^{k}~_{mnr}\cdot g^{rl}\cdot u_{\overline{l}}\cdot u_{j}\text{ \ ,}%
\end{align*}
\begin{align*}
R^{k}~_{mnr}\cdot g^{rl}\cdot u_{\overline{l}}  &  =R^{k}~_{mnr}\cdot
u^{r}\text{ \ ,}\\
u_{\overline{k}}\cdot R^{k}~_{mnr}  &  =g_{\overline{k}\overline{s}}\cdot
u^{s}\cdot R^{k}~_{mnr}\text{ \ ,}%
\end{align*}
we have%

\begin{align}
h_{i\overline{k}}\cdot R^{k}~_{mnr}\cdot g^{rl}\cdot h_{\overline{l}j}  &
=g_{i\overline{k}}\cdot R^{k}~_{mnr}\cdot g^{rl}\cdot g_{\overline{l}%
j}-\nonumber\\
&  -\frac1e\cdot u_{i}\cdot g_{\overline{k}\overline{s}}\cdot u^{s}\cdot
R^{k}~_{mnr}\cdot g^{rl}\cdot g_{\overline{l}j}-\nonumber\\
&  -\frac1e\cdot g_{i\overline{k}}\cdot R^{k}~_{mnr}\cdot g^{rl}\cdot
g_{\overline{l}\overline{s}}\cdot u^{s}\cdot u_{j}+\label{(3.8)}\\
&  +\frac1{e^{2}}\cdot u_{i}\cdot g_{\overline{k}\overline{s}}\cdot u^{s}\cdot
R^{k}~_{mnr}\cdot g^{rl}\cdot g_{\overline{l}\overline{q}}\cdot u^{q}\cdot
u_{j}\text{ .}\nonumber
\end{align}

\textit{Special case:} $V_{n}$-space: $S:=C$.
\begin{equation}
M_{ij}=R_{imnj}\cdot u^{m}\cdot u^{n}\text{ \ \ \ \ .} \label{3.9}%
\end{equation}
\begin{align}
_{s}D_{ij}  &  =~_{s}M_{ij}=M_{ij}-\frac{1}{n-1}\cdot I\cdot h_{ij}%
=\nonumber\\
&  =R_{imnj}\cdot u^{m}\cdot u^{n}-\frac{1}{n-1}\cdot R_{mn}\cdot u^{m}\cdot
u^{n}\cdot h_{ij\text{ }}\text{,}\nonumber\\
_{s}D_{ij}  &  =(R_{imnj}-\frac{1}{n-1}\cdot R_{mn}\cdot h_{ij\text{ }})\cdot
u^{m}\cdot u^{n}=~_{s}M_{ij}\text{ ,}\label{3.10}\\
U  &  =I=R_{ij}\cdot u^{i}\cdot u^{j}\text{ .}\nonumber
\end{align}

In a $V_{n}$-space the components $a_{z}^{i}$ of the centrifugal (centripetal)
acceleration $a_{z}$ have the form
\begin{align}
a_{z}^{i}  &  =(\frac{1}{n-1}\cdot U+\frac{_{s}D_{jk}\cdot\xi_{\perp}^{j}%
\cdot\xi_{\perp}^{k}}{g_{rs}\cdot\xi_{\perp}^{r}\cdot\xi_{\perp}^{s}})\cdot
\xi_{\perp}^{i}=\label{3.11}\\
&  =[\frac{1}{n-1}\cdot R_{mn}\cdot u^{m}\cdot u^{n}+\nonumber\\
&  +\frac{1}{g_{rs}\cdot\xi_{\perp}^{r}\cdot\xi_{\perp}^{s}}\cdot
(R_{jmnk}-\frac{1}{n-1}\cdot R_{mn}\cdot h_{jk})\cdot u^{m}\cdot u^{n}\cdot
\xi_{\perp}^{j}\cdot\xi_{\perp}^{k}]\cdot\xi_{\perp}^{i}\text{ .}\nonumber
\end{align}

3. If the Einstein equations in vacuum without cosmological term $(\lambda
_{0}=0)$ are valid, i.e. if
\begin{equation}
R_{ij}=0\text{ \ , \ \ }\lambda_{0}=\text{const.}=0\text{ \ , \ \ \ \ }%
n=4\text{ ,} \label{3.12}%
\end{equation}
are fulfilled then
\begin{align}
a_{z}^{i}  &  =\frac{1}{g_{rs}\cdot\xi_{\perp}^{r}\cdot\xi_{\perp}^{s}}\cdot
R_{jmnk}\cdot u^{m}\cdot u^{n}\cdot\xi_{\perp}^{j}\cdot\xi_{\perp}^{k}\cdot
\xi_{\perp}^{i}=\label{3.13}\\
&  =R_{jmnk}\cdot u^{m}\cdot u^{n}\cdot n^{j}\cdot n^{k}\cdot\xi_{\perp}%
^{i}=\mathbf{g}\cdot\xi_{\perp}^{i}\text{ ,} \label{3.14}%
\end{align}
where
\begin{equation}
n^{j}=\frac{\xi_{\perp}^{j}}{\sqrt{\mid g_{rs}\cdot\xi_{\perp}^{r}\cdot
\xi_{\perp}^{s}\mid}}\text{ ,} \label{3.15}%
\end{equation}
\begin{equation}
\mathbf{g=}R_{jmnk}\cdot u^{m}\cdot u^{n}\cdot n^{j}\cdot n^{k}\text{ \ .}
\label{3.16}%
\end{equation}

If $a_{z}$ is a centripetal acceleration interpreted as gravitational
acceleration for a free spinless test particles moving in an external
gravitational field $(R_{ij}=0)$ then the condition $\mathbf{g<~}0$ should be
valid if \ the centripetal acceleration is directed to the particle. If the
centripetal acceleration is directed to the gravitational source (in the
direction $\xi_{\perp}$) then $\mathbf{g>~}0$.

From a more general point of view as that in the ETG, a gravitational theory
could be worked out in a $(\overline{L}_{n},g)$-space where $a_{z}$ could also
be interpreted as gravitational acceleration of mass elements or particles
generating a gravitational field by themselves and caused by their motions in space-time.

4. If we consider a frame of reference in which a mass element (particle) is
at rest then $u^{i}=g_{4}^{i}\cdot u^{4}$ and $\xi_{\perp}^{i}=g_{a}^{i}%
\cdot\xi_{\perp}^{a}:=g_{1}^{i}\cdot\xi_{\perp}^{1}$. The centrifugal
(centripetal) acceleration $a_{z}^{i}$ could be written in the form
\begin{equation}
a_{z}^{i}=R_{1441}\cdot u^{4}\cdot u^{4}\cdot n^{1}\cdot n^{1}\cdot\xi_{\perp
}^{i}=R_{1441}\cdot(u^{4})^{2}\cdot(n^{1})^{2}\cdot\xi_{\perp}^{i}\text{ \ .}
\label{3.17}%
\end{equation}

For the Schwarzschild metric
\begin{align}
ds^{2}  &  =\frac{dr^{2}}{1-\frac{r_{g}}r}+r^{2}\cdot(d\theta^{2}+\sin
^{2}\theta\cdot d\varphi^{2})-(1-\frac{r_{g}}r)\cdot(dx^{4})^{2}\text{ \ ,
}\label{3.18}\\
\text{\ }r_{g}  &  =\frac{2\cdot k\cdot M_{0}}{c^{2}}\text{ ,} \label{3.18a}%
\end{align}
the component $R_{1441}=g_{11}\cdot R^{1}~_{441}$ of the curvature tensor has
the form \cite{Stephani}
\begin{equation}
R_{1441}=-g_{11}\cdot(\Gamma_{44,1}^{1}-\Gamma_{14,4}^{1}+\Gamma_{11}^{1}%
\cdot\Gamma_{44}^{1}+\Gamma_{14}^{1}\cdot\Gamma_{44}^{4}-\Gamma_{14}^{1}%
\cdot\Gamma_{14}^{1}-\Gamma_{44}^{1}\cdot\Gamma_{41}^{4})\text{ .}
\label{3.19}%
\end{equation}

After introducing in the last expression the explicit form of the metric and
of the Christoffel symbols $\Gamma_{jk}^{i}$, it follows for $R_{1441}$%
\begin{equation}
R_{1441}=\frac{r_{g}}{r^{3}}\text{ \ .} \label{3.20}%
\end{equation}

Then
\begin{equation}
a_{z}^{i}=\frac{r_{g}}{r^{3}}\cdot(u^{4})^{2}\cdot(n^{1})^{2}\cdot\xi_{\perp
}^{i}\text{ .} \label{3.21}%
\end{equation}

If the co-ordinate time $t=x^{4}/c$ is chosen as equal to the proper time
$\tau$ of the particle, i.e. if $t=\tau$ then
\begin{align}
u^{4}  &  =\frac{dx^{4}}{d\tau}=c\cdot\frac{d\tau}{d\tau}=c\text{ ,
\ \ \ \ \ }n^{1}=1\text{ \ \ , \ \ \ \ \ \ }\xi_{\perp}^{i}=g_{1}^{i}\cdot
\xi_{\perp}^{1}\text{ \ ,}\label{3.22}\\
a_{z}^{i}  &  =\frac{r_{g}}{r^{3}}\cdot c^{2}\cdot g_{1}^{i}\cdot\xi_{\perp
}^{1}\text{ , \ \ \ \ \ }\frac{r_{g}}{r^{3}}\cdot c^{2}>0\text{ \ \ .}
\label{3.23}%
\end{align}

For the centrifugal (centripetal) \ acceleration we obtain
\begin{equation}
a_{z}^{1}=\frac{2\cdot k\cdot M_{0}}{r^{3}}\cdot\xi_{\perp}^{1}\text{ \ \ ,}
\label{3.24}%
\end{equation}
which is exactly the relative gravitational acceleration between two mass
elements (particles) with co-ordinates $x_{c1}=r$ and $x_{c2}=r+\xi_{\perp
}^{1}$ \cite{Manoff-9}.

Therefore, the centrifugal (centripetal) acceleration could be used for
working out of a theory of gravitation in a space with affine connections and
metrics as this has been done in the Einstein theory of gravitation.

\section{Conclusions}

In the present paper the notions of (relative) centrifugal (centripetal) and
(relative) Coriolis' velocities and accelerations are introduced and
considered in spaces with affine connections and metrics as velocities and
accelerations of flows of mass elements (particles) moving in space-time. It
is shown that these types of velocities and accelerations are generated by the
relative motions between the mass elements. The centrifugal (centripetal)
velocity is found to be in connection with the Hubble law in spaces with
affine connections and metrics. [The relations between null vector fields and
the Hubble and the Doppler effects will be considered elsewhere
\cite{Manoff-16}.] The accelerations are closely related to the kinematic
characteristics of the relative velocity and relative acceleration. The
centrifugal (centripetal) acceleration could be interpreted as gravitational
acceleration as it has been done in the Einstein theory of gravitation. This
fact could be used as a basis for working out of new gravitational theories in
spaces with affine connections and metrics.


\begin{thebibliography}{99}                                                                                               %


\bibitem {Manoff-0}S. Manoff: "Mechanics of continuous media in $(\overline
{L}_{n},g)$-spaces.\ 1. Introduction and mathematical tools"
(\textit{Preprint} gr-qc /0203016), 2002.

\bibitem {Manoff-1}S. Manoff: "Mechanics of continuous media in $(\overline
{L}_{n},g)$-spaces. 2. Relative velocity and deformations" (\textit{Preprint}
gr-qc /0203017), 2002.

\bibitem {Manoff-3}S. Manoff: "Mechanics of continuous media in $(\overline
{L}_{n},g)$-spaces. 3. Relative accelerations" (\textit{Preprint} gr-qc
/0204003), 2002.

\bibitem {Manoff-4}S. Manoff: "Mechanics of continuous media in $(\overline
{L}_{n},g)$-spaces. 4. Stress (tension) tensor". (\textit{Preprint} gr-qc
/0204004), 2002.

\bibitem {Stephani}H. Stephani: "\textit{Allgemeine Relativitaetstheorie".}
VEB Deutscher Verlag d. Wissenschaften, Berlin, 1977, pp. 75-76.

\bibitem {Ehlers}J. Ehlers: "Beitraege zur relativistischen Mechanik
kontinuierlicher Medien". \textit{Abhandlungen d. Mainzer Akademie d.
Wissenschaften, Math.-Naturwiss. Kl}. Nr. \textbf{11 }(1961), 792-837.

\bibitem {Lammerzahl}Cl. Laemmerzahl: "A Characterisation of the Weylian
structure of Space-Time by Means of Low Velocity Tests" (\textit{Preprint}
gr-qc /0103047), 2001.

\bibitem {Bishop}R. L. Bishop, S. I. Goldberg. "\textit{Tensor Analysis on
Manifolds".} The Macmillan Company, New York, 1968.

\bibitem {Manoff-5}S. Manoff: "Kinematics of vector fields". \textit{Complex
Structures and Vector Fields} ed St Dimiev and K Sekigawa. World Scientific,
Singapore, 1995, pp 61-113.

\bibitem {Manoff-6}S. Manoff: "Spaces with contravariant and covariant affine
connections and metrics". \textit{Phys. Part. Nuclei} \textbf{30} (1999) 5,
527-49 (Russian edn: \textbf{30} (1999) 5, 1211-69).

\bibitem {Manoff-7}S. Manoff: "\textit{Geometry and Mechanics in Different
Models of Space-Time}: \textit{Geometry and Kinematics"}. Nova Science
Publishers, New York, 2002.

\bibitem {Manoff-7a}S. Manoff: "\textit{Geometry and Mechanics in Different
Models of Space-Time}: \textit{Dynamics and Applications"}. Nova Science
Publishers, New York, 2002.

\bibitem {Manoff-8}S. Manoff: "Einstein's theory of gravitation as a
Lagrangian theory for tensor fields". \textit{Intern. J. Mod. Phys}. \textbf{A
13} (1998) 12, 1941-67.

\bibitem {Manoff-9}S. Manoff: "About the motion of test particles in an
external gravitational field". \textit{Exp. Technik der Physik} \textbf{24}
(1976) 5, 425-431 (in German).

\bibitem {Misner}Ch. W. Misner, K. S. Thorne, J. A. Wheeler:
"\textit{Gravitation". }W. H. Freeman and Company, San Francisco, 1973.
(Russian translation 1977 Mir, Moscow, Vol 1, Vol. 2, Vol. 3).

\bibitem {Manoff-15}S. Manoff: "Flows and particles with shear-free and
expansion-free velocities in $(L_{n},g)$- and Weyl's spaces". \textit{Class.
Quantum Grav.} \textbf{19} (2002 ) 16, 4377-98 (\textit{Preprint} gr-qc/0207060).

\bibitem {Manoff-16}S. Manoff: "Centrifugal (centripetal), Coriolis'
velocities, accelerations and Hubble's law in spaces with affine connections
and metrics" (\textit{Preprint} gr-qc/0212038), 2002.
\end{thebibliography}
\end{document}